\pdfoutput=1
\newif\ifplainstyle
\plainstyletrue
\plainstylefalse

\newif\ifjhepstyle
\jhepstyletrue

\newif\ifprstyle
\prstyletrue
\prstylefalse

\ifprstyle
	\documentclass[twocolumn,nofootinbib]{revtex4-1}
\else
	\documentclass[11pt,a4paper]{article}
\fi

\ifjhepstyle
	\usepackage{jheppub}
	\usepackage{amsfonts}
	\usepackage{verbatim}
	\usepackage{float}
	\usepackage{color}

	\usepackage{array}
	\newcolumntype{C}[1]{>{\centering\arraybackslash$}p{#1}<{$}}
	\makeatletter
	\def\@fpheader{\phantom{:-)}}
	\makeatother
\else	
 	\ifprstyle
		\usepackage{verbatim}
		\usepackage{amsmath,amsfonts,amssymb}
		\usepackage[colorlinks=true
                	,urlcolor=blue
                	,anchorcolor=blue
                	,citecolor=blue
                	,filecolor=blue
                	,linkcolor=blue
                	,menucolor=blue
                	]{hyperref}
	\else
            	\usepackage{verbatim}
            	\usepackage{cite}
            	\usepackage{setspace}
            	\usepackage[top=2.5cm, bottom=2.75cm, left=2.5cm, right=2.5cm]{geometry}
            	\usepackage{amsmath,amsfonts,amssymb}
            	\usepackage[colorlinks=true
            	,urlcolor=blue
            	,anchorcolor=blue
            	,citecolor=blue
            	,filecolor=blue
            	,linkcolor=blue
            	,menucolor=blue
            	,linktoc=page
            	]{hyperref}
            	
            	\usepackage{float}
            	\restylefloat{table}
            	
            	\numberwithin{equation}{section}
            	
	\fi
\fi

\usepackage{subfig}

\usepackage{etoolbox}

\makeatletter
\def\hlinewd#1{%
\noalign{\ifnum0=`}\fi\hrule \@height #1 %
\futurelet\reserved@a\@xhline}
\makeatother

\newcolumntype{?}[1]{!{\vrule width #1}}

\allowdisplaybreaks

\usepackage{multirow}

\usepackage[normalem]{ulem}
\usepackage{mathtools}
\usepackage{transparent}
\usepackage{bm}
\usepackage{enumitem}

\usepackage[textsize=scriptsize,textwidth=2.5cm]{todonotes}

\makeatletter
\let\save@mathaccent\mathaccent
\newcommand*\if@single[3]{%
  \setbox0\hbox{${\mathaccent"0362{#1}}^H$}%
  \setbox2\hbox{${\mathaccent"0362{\kern0pt#1}}^H$}%
  \ifdim\ht0=\ht2 #3\else #2\fi
  }
\newcommand*\rel@kern[1]{\kern#1\dimexpr\macc@kerna}
\newcommand*\widebar[1]{\@ifnextchar^{{\wide@bar{#1}{0}}}{\wide@bar{#1}{1}}}
\newcommand*\wide@bar[2]{\if@single{#1}{\wide@bar@{#1}{#2}{1}}{\wide@bar@{#1}{#2}{2}}}
\newcommand*\wide@bar@[3]{%
  \begingroup
  \def\mathaccent##1##2{%
    \let\mathaccent\save@mathaccent
    \if#32 \let\macc@nucleus\first@char \fi
    \setbox\z@\hbox{$\macc@style{\macc@nucleus}_{}$}%
    \setbox\tw@\hbox{$\macc@style{\macc@nucleus}{}_{}$}%
    \dimen@\wd\tw@
    \advance\dimen@-\wd\z@
    \divide\dimen@ 3
    \@tempdima\wd\tw@
    \advance\@tempdima-\scriptspace
    \divide\@tempdima 10
    \advance\dimen@-\@tempdima
    \ifdim\dimen@>\z@ \dimen@0pt\fi
    \rel@kern{0.6}\kern-\dimen@
    \if#31
      \overline{\rel@kern{-0.6}\kern\dimen@\macc@nucleus\rel@kern{0.4}\kern\dimen@}%
      \advance\dimen@0.4\dimexpr\macc@kerna
      \let\final@kern#2%
      \ifdim\dimen@<\z@ \let\final@kern1\fi
      \if\final@kern1 \kern-\dimen@\fi
    \else
      \overline{\rel@kern{-0.6}\kern\dimen@#1}%
    \fi
  }%
  \macc@depth\@ne
  \let\math@bgroup\@empty \let\math@egroup\macc@set@skewchar
  \mathsurround\z@ \frozen@everymath{\mathgroup\macc@group\relax}%
  \macc@set@skewchar\relax
  \let\mathaccentV\macc@nested@a
  \if#31
    \macc@nested@a\relax111{#1}%
  \else
    \def\gobble@till@marker##1\endmarker{}%
    \futurelet\first@char\gobble@till@marker#1\endmarker
    \ifcat\noexpand\first@char A\else
      \def\first@char{}%
    \fi
    \macc@nested@a\relax111{\first@char}%
  \fi
  \endgroup
}
\makeatother
\newcommand{\ThisIsTheTitle}{Holographic Complexity of Rotating Quantum Black Holes

} 
\newcommand{\ThisIsAuthorOne}{Bin Chen,}
\newcommand{\ThisIsEmailOne}{bchen01@pku.edu.cn}
\newcommand{\ThisIsAuthorTwo}{Yuefeng Liu}
\newcommand{\ThisIsEmailTwo}{yfliu0905@pku.edu.cn}
\newcommand{\ThisIsAuthorThree}{and Boyang Yu}
\newcommand{\ThisIsEmailThree}{yuby21@pku.edu.cn}
 
\newcommand{\TheseAreTheKeywords}{}

\newcommand{\ThisIsTheAbstract}{
We study  holographic complexity for the rotating quantum BTZ black holes (quBTZ), the BTZ black holes with corrections from bulk quantum fields. Using double holography, the combined system of backreacted rotating BTZ black holes with conformal matters,  can be holographically described by the rotating AdS$_4$ C-metric with the BTZ black hole living on a codimension-1 brane. We investigate both volume complexity and action complexity of rotating quBTZ, and pay special attention to their late-time behaviors. When the mass of BTZ black hole is not very small and the rotation is not very slow,  we show that the late-time rates of the volume complexity and the action complexity agree with each other up to a factor $2$ and reduce to the ones of BTZ at the leading classical order, and they both receive subleading quantum corrections. 
For the volume complexity, the leading quantum correction comes from the backreaction of comformal matter on the geometry, similar to the static quBTZ case. For the action complexity, unlike the static case, the Wheeler-de Witt (WdW) patch in computing the action complexity for the rotating black hole does not touch the black hole singularity such that the leading order result is in good match with  the one  of classical BTZ.   However, when the mass of BTZ black hole is small or the rotation parameter $a$ is small, the quantum correction to the action complexity could be significant such that the late-time slope of the action complexity of quBTZ deviates very much from the one of classical BTZ. Remarkably, we notice that the nonrotating limit $a\to 0$ is singular and does not lead to the late-time slope of the action complexity for non-rotating quantum BTZ black hole. The similar phenomenon happens for higher dimensional rotating black holes.

}

\ifjhepstyle
\title{\ThisIsTheTitle}

\author[ a, b,c ]{\ThisIsAuthorOne}
\author[ a ]{\ThisIsAuthorTwo}
\author[ b ]{\ThisIsAuthorThree}

\affiliation[a]{Department of Physics, Peking University, No.5 Yiheyuan Rd, Beijing 100871, P.R. China}
\affiliation[b]{Center for High Energy Physics, Peking University, No.5 Yiheyuan Rd, Beijing 100871, P.
R. China}

\affiliation[c]{Collaborative Innovation Center of Quantum Matter, No.5 Yiheyuan Rd, Beijing 100871,
P. R. China}

\emailAdd{\ThisIsEmailOne}
\emailAdd{\ThisIsEmailTwo}
\emailAdd{\ThisIsEmailThree}

\abstract{\ThisIsTheAbstract} 

\fi

\begin{document}

\ifjhepstyle
\maketitle
\flushbottom
\fi

\long\def\symfootnote[#1]#2{\begingroup%
\def\thefootnote{\fnsymbol{footnote}}\footnote[#1]{#2}\endgroup} 

\def\rednote#1{{\color{red} #1}}
\def\bluenote#1{{\color{blue} #1}}
\def\magnote#1{{\color{magenta} #1}}
\def\rout#1{{\color{red} \sout{#1}}}

\newcommand{\BC}[2]{\textcolor{magenta}{#1}\todo[color=green]{\scriptsize{BC: #2}}}
\newcommand{\LY}[2]{\textcolor{red}{#1}\todo[color=yellow]{\scriptsize{LY: #2}}}
\newcommand{\by}[2]{\textcolor{blue}{#1}\todo[color=yellow]{\scriptsize{BY: #2}}}

\def\({\left (}
\def\){\right )}
\def\lb{\left [}
\def\rb{\right ]}
\def\lB{\left \{}
\def\rB{\right \}}

\def\Int#1#2{\int \textrm{d}^{#1} x \sqrt{|#2|}}
\def\Bra#1{\left\langle#1\right|} 
\def\Ket#1{\left|#1\right\rangle}
\def\BraKet#1#2{\left\langle#1|#2\right\rangle} 
\def\Vev#1{\left\langle#1\right\rangle}
\def\Vevm#1{\left\langle \Phi |#1| \Phi \right\rangle}\def\bbox{\bar{\Box}}
\def\til#1{\tilde{#1}}
\def\wtil#1{\widetilde{#1}}
\def\ph#1{\phantom{#1}}

\def\ra{\rightarrow}
\def\la{\leftarrow}
\def\lra{\leftrightarrow}
\def\p{\partial}
\def\barp{\bar{\partial}}
\def\diff{\mathrm{d}}

\def\sinh{\mathrm{sinh}}
\def\cosh{\mathrm{cosh}}
\def\tanh{\mathrm{tanh}}
\def\coth{\mathrm{coth}}
\def\sech{\mathrm{sech}}
\def\csch{\mathrm{csch}}

\def\a{\alpha}
\def\b{\beta}
\def\g{\gamma}
\def\d{\delta}
\def\e{\epsilon}
\def\ve{\varepsilon}
\def\k{\kappa}
\def\l{\lambda}
\def\n{\nabla}
\def\om{\omega}
\def\s{\sigma}
\def\t{\theta}
\def\z{\zeta}
\def\vp{\varphi}

\def\ss{\Sigma}
\def\dd{\Delta}
\def\GG{\Gamma}
\def\LL{\Lambda}
\def\tt{\Theta}

\def\A{{\cal A}}
\def\B{{\cal B}}
\def\C{{\cal C}}
\def\cE{{\cal E}}
\def\D{{\cal D}}
\def\F{{\cal F}}
\def\H{{\cal H}}
\def\I{{\cal I}}
\def\J{{\cal J}}
\def\K{{\cal K}}
\def\L{{\cal L}}
\def\M{{\cal M}}
\def\N{{\cal N}}
\def\O{{\cal O}}
\def\Q{{\cal Q}}
\def\P{{\cal P}}
\def\cS{{\cal S}}
\def\T{{\cal T}}
\def\W{{\cal W}}
\def\X{{\cal X}}
\def\Z{{\cal Z}}

\def\mfa{\mathfrak{a}}
\def\mfb{\mathfrak{b}}
\def\mfc{\mathfrak{c}}
\def\mfd{\mathfrak{d}}

\def\we{\wedge}
\def\re{\textrm{Re}}

\def\tilw{\tilde{w}}
\def\tile{\tilde{e}}

\def\tilL{\tilde{L}}
\def\tilJ{\tilde{J}}

\def\zz{\bar z}
\def\xx{\bar x}
\def\yy{\bar y}
\def\xp{x^{+}}
\def\xm{x^{-}}

\def\bp{\bar{\p}}
\def\note#1{{\color{red}#1}}
\def\notebf#1{{\bf\color{red}#1}}

\def\vac{\text{vac}}

\def\VirU1{Vir \times U(1)}
\def\VirSL2R{\mathrm{Vir}\otimes\widehat{\mathrm{SL}}(2,\mathbb{R})}
\def\U1{U(1)}
\def\u1{U(1)}
\def\SL2R{\widehat{\mathrm{SL}}(2,\mathbb{R})}
\def\sl2r{\mathrm{SL}(2,\mathbb{R})}

\def\RR{\mathbb{R}}

\def\tr{\mathrm{Tr}}
\def\bnabla{\overline{\nabla}}

\def\sint{\int_{\ss}}
\def\dsint{\int_{\p\ss}}
\def\hint{\int_{H}}

\def\sym{\textrm{Sym}}
\def\symTT{\textrm{Sym}^N \mathcal M_\mu}
\def\symCFT{\textrm{Sym}^N \mathcal M_0}

\newcommand{\eq}[1]{\begin{align}#1\end{align}}
\newcommand{\eqst}[1]{\begin{align*}#1\end{align*}}
\newcommand{\eqsp}[1]{\begin{equation}\begin{split}#1\end{split}\end{equation}}

\newcommand{\absq}[1]{{\textstyle\sqrt{\left |#1\right |}}}



\newcommand{\pp}{{\partial_+}{}}
\newcommand{\pmm}{{\partial_-}{}}

\newcommand{\be}{\begin{equation}}
\newcommand{\ee}{\end{equation}}
\newcommand{\bea}{\begin{eqnarray}}
\newcommand{\eea}{\end{eqnarray}}
\newcommand{\bb}{\mathbb}
\newcommand{\ba}{\begin{aligned}}
\newcommand{\ea}{\end{aligned}}
\newcommand{\me}{\mathcal{E}}
\newcommand{\yf}{\textcolor{cyan}}
\newcommand{\BY}{\textcolor{blue}}


\ifprstyle
\title{\ThisIsTheTitle}

\author{\ThisIsAuthorOne}
\email{\ThisIsEmailOne}

\author{\ThisIsAuthorTwo}
\email{\ThisIsEmailTwo}

\affiliation{\ThisIsTheAffiliation}


\begin{abstract}
\ThisIsTheAbstract
\end{abstract}


\maketitle

\fi
\ifplainstyle
\begin{titlepage}
\begin{center}

\ph{.}

\vskip 4 cm

{\Large \bf \ThisIsTheTitle}

\vskip 1 cm

\renewcommand*{\thefootnote}{\fnsymbol{footnote}}

{{\ThisIsAuthorOne}\footnote{\ThisIsEmailOne} } and {\ThisIsAuthorTwo}\footnote{\ThisIsEmailTwo}

\renewcommand*{\thefootnote}{\arabic{footnote}}

\setcounter{footnote}{0}

\vskip .75 cm


\end{center}

\vskip 1.25 cm
\date{}


\end{titlepage}

\newpage

\fi

\ifplainstyle
\tableofcontents
\noindent\hrulefill
\bigskip
\fi

\section{Introduction}

Within the framework of gauge/gravity duality \cite{Maldacena:1997re,Aharony:1999ti,Witten:1998qj}, a flux of notions from quantum information theory has been introduced into the study of  hidden structure of quantum spacetime, and has led to  prominent and intriguing insights in recent years. A prime example  is the  holographic entanglement entropy proposed by Ryu-Takayanagi\cite{Ryu:2006bv,Hubeny:2007xt}, manifesting the idea that geometry is emergent. However, it has been recognized that this measure alone is not sufficient to fully describe the rich structures of the bulk spacetime, especially the mysterious interior of black hole. To address this problem,  the quantum computation complexity was introduced into the play\cite{Susskind:2014moa}. The complexity in quantum information  quantifies how hard it is to prepare a particular state of interest by applying a series of elementary gates to
a reference state. To compute holographic complexity at a given time, there are two different proposals: one is  the ``complexity=volume" (CV) proposal \cite{Susskind:2014rva,Stanford:2014jda} and  the other is the ``complexity=action" (CA) proposal \cite{Brown:2015bva,Brown:2015lvg}.

A classic example for discussing holographic complexity is the external two-sided black hole, where the bulk geometry is connected by the Einstein-Rosen bridge. In the field theory, the state  dual to this geometry is the thermofield double state,
\be 
| TFD(t_{L},t_{R})\rangle = Z^{-1/2} \sum_{n} e^{-E_{n}/(2T)} e^{-i E_{n}(t_{L}+t_{R})} |E_{n} \rangle_{L}  | E_{n} \rangle_{R}.
\ee
Similarly for the rotating black hole, the dual state is the TFD state prepared with the deformed Hamiltonian $\beta (H+\Omega J)$.  The volume complexity (VC) conjecture states that the complexity of state $|TFD\rangle$ is dual to the extremal/maximal volume of bulk co-dimension-one hypersurface anchored at fixed $t_L$ and $t_R$ boundary time slice,
\be \label{def:CV}
   C_{V}(|TFD\rangle)= \textbf{Max} \( \frac{\mbox{Vol}(\Sigma)}{G L \hbar} \),
\ee
where $L$ is a CFT length scale related to AdS radius $L\sim l_{AdS}$. The action complexity (AC) conjecture states that the complexity is captured by  the on-shell gravitational action on a region of spacetime known as the Wheeler-DeWitt (WdW)
patch anchored at $t_L$ and $t_R$ on asymptotic boundaries,
\be 
 C_{A}=\frac{I_{WdW}}{\pi \hbar}.
\ee
While the above two proposals do not yield  exactly the same results, nevertheless they share similar qualitative features. For example, at late-time regime, both complexities for various kinds of spacetimes are expected to be proportional to the thermodynamic quantity $TS$
\be \frac{d C_{V,A}}{dt}\bigg|_{t\gg \beta} \sim T S,\ee
up to $\mathcal{O}(1)$ constant coefficients depending on the dimension and the details of spacetimes. 

There have been intensive investigations on the holographic complexity from various points of view. Here we would like to discuss the quantum corrections from the bulk fields to the complexity. The quantum corrections from the bulk fields to the holographic complexity is of order $\mathcal{O}(G^0)$, similar to the quantum correction to the holographic entanglement entropy\cite{Barrella:2013wja,Faulkner:2013ana,Chen:2013kpa,Chen:2013dxa,Chen:2014unl,Chen:2015kua}. These quantum corrections have been studied in 2D dilaton gravity\cite{Yang:2018gdb,Lin:2019qwu,Schneiderbauer:2019anh,Schneiderbauer:2020isp,Iliesiu:2021ari}. They have also been investigated for the quantum static BTZ black hole in the framework of double holography \cite{Emparan:2021hyr}. In the present work, we would like to generalize the study to the quantum rotating BTZ black hole, whose double holographic dual is the  AdS$_4$ rotating C-metric.

Our main motivation is to figure out if there exists any universal feature in holographic complexity independent of the specific state. In \cite{Emparan:2021hyr}, it was showed that the action complexity of static quBTZ is very different from the one of classical BTZ, possibly due to the large quantum correction near the spacetime singularity. It would be interesting to see if this happens as well in the rotating case. According to our limit knowledge, there were only several interesting works \cite{AlBalushi:2020heq, Bernamonti:2021jyu} discussing the effects of rotation in the context of holographic complexity at the classic level \footnote{See also the interesting works \cite{Auzzi:2018pbc,Auzzi:2018zdu}, which contain the results of rotating BTZ when the warping number equals to 1. We thank the authors for pointing out these to us.}. It will be nice to study the effect of rotation on the quantum correction as well. 

Technically, the investigation of the complexity for a rotational spacetime is more challenging. For example, in computing the action complexity of  4d Kerr-(A)ds spacetimes \cite{AlBalushi:2019obu}, one has to analyze the caustic of the WdW patch, which is a formidable task. For the AdS$_4$ rotating C-metric, which describes the rotating quBTZ spacetime, we manage to show that the caustics is absent about the regularized WdW patch as well, after detailed analysis. 

In order to study the quantum effects from the conformal matter, we try to do a semi-classical expansion of the complexities. This is feasible for the volume complexity, but is tricky for the action complexity.
In this work, we mainly discuss the late-time growth rate of the holographic complexity. 
For the volume complexity, its late-time growth rate can reduce to the one of classical BTZ at the leading order, and it receives the quantum correction from the backreaction of the CFT matter to the background geometry. For the action complexity, if both the mass and the spin of the BTZ black hole are not small, its late-time growth rate can be decomposed into the classical part and the quantum-corrected part, with the classical part being the same as the one of classical BTZ. 
More remarkably, when the mass of BTZ is small or the spin is small, the late-time growth rate of action complexity for rotating quBTZ could receive significant quantum correction such that it shows large discrepancy from the one of classical BTZ.

One remarkable point from our study is the singular behavior of the nonrotating limit. We show that under this limit, the late-time slope of the action complexity of rotating quBTZ does not reproduce the one of the static case. We analyze the underlying reason behind this singular behavior, and find it exists in the higher dimensional rotating black holes as well.

The remaining parts of this paper are organized as follows. In section 2, we introduce our setup, the double holography,  and discuss the properties of the AdS$_4$ rotating C-metric,  especially its thermodynamics. In section 3, we compute the volume complexity and identify the leading quantum-correction from the three-dimensional effective theory on the brane. In section 4, we study the action complexity and compute its late-time growth rate. We find that  the leading-order  result matches the one of volume complexity in most of parameter space. In the case that the mass or the spin of BTZ black hole is small,  the  quantum correction could be significant. We notice the discontinuity of the late-time derivative of the action complexity in the non-rotating limit, and find that this singular behavior exists  for higher dimensional black holes as well. We end with some discussions in section 5. Some technical details, including the proof of absence of caustics, the nonrotating limit in the case of higher dimensional Kerr-AdS black hole, the chargeless limit in the case of RN-AdS black holes, are collected in the appendices.


\section{Double holography and quantum rotating BTZ}

\begin{figure}[h]
\centering
\includegraphics[scale=0.8]{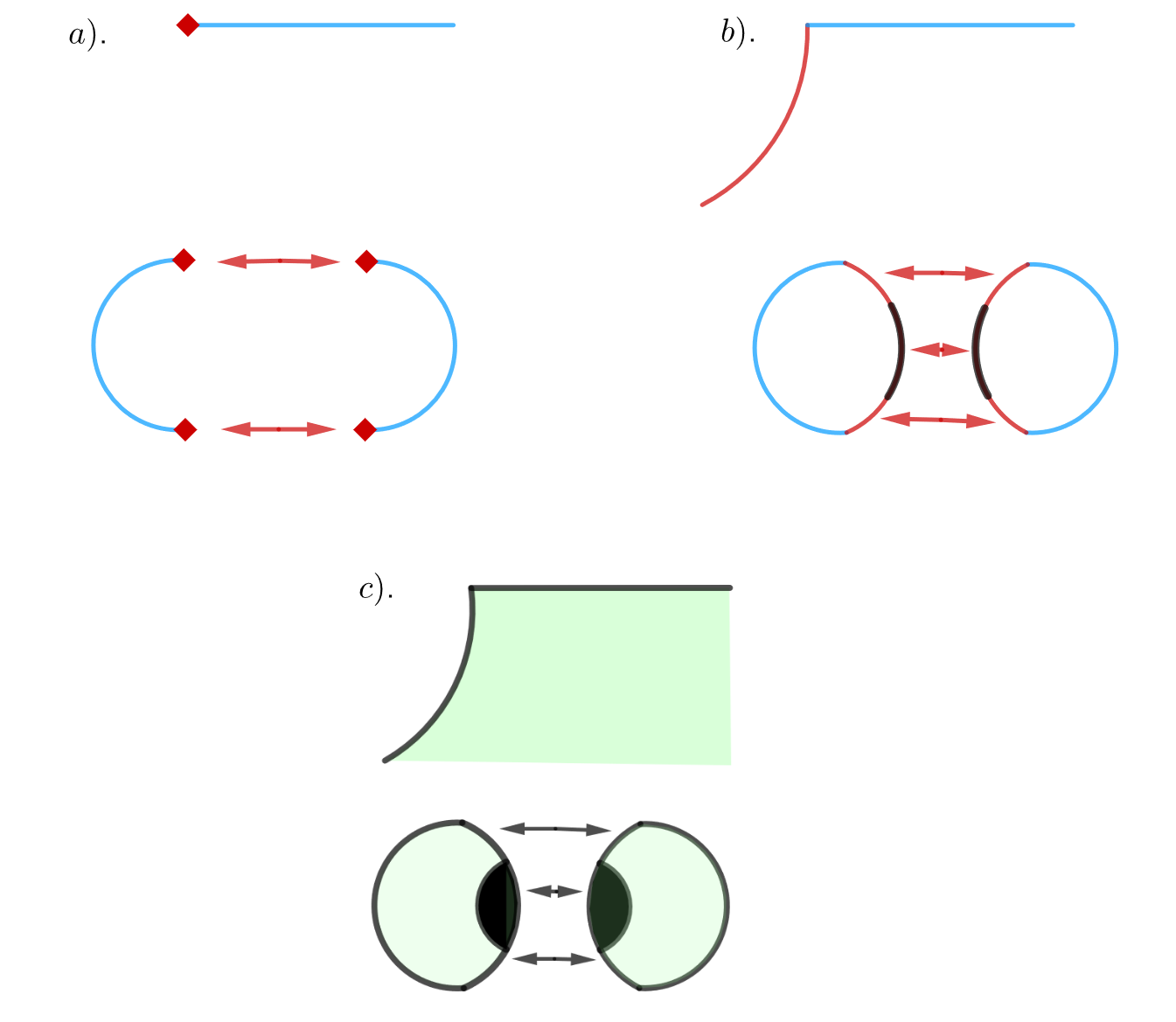}
\caption{Carton pictures for a spatial slice in the holographic braneworld model. In all three figures, the top rows show the three different perspectives of the model, while the bottom rows display the two-sided setup. a). Pure boundary field perspective, where two
identified 2d conformal defects (red diamonds) coupled to two 3d bath CFTs (blue lines) respectively; b). Brane perspective, where the defect is replaced with an effective 3d gravitational theory on the branes (red arcs) couple to the 3d bath CFTs (blue lines) through transparent
boundary conditions; c). Pure bulk perspective, where the left (right) bulk patch (green region) consists of a rotating black hole (black region) intersecting
a backreacted Planck brane (black arc). The patches then are glued together along the brane via
Israel’s junction conditions.  }
\label{fig:intor}
\end{figure}

We are going to work in the holographic braneworld model, or called double holography, to investigate the quantum corrections to the holographic complexity. The basic idea  is to consider two possible dual descriptions of a codimension-1 conformal defect in a CFT. One dual picture is to describe the defect by a AdS gravity with conformal matter which  asymptotically couples to the bulk CFT.  The other dual picture, which is referred to as double holography, is to describe the whole system in a higher dimensional AdS gravity with a Planckian brane\cite{Karch:2000ct,DeWolfe:2001pq}. If the whole system is in a thermal state, in the first picture the defect could be described by a semiclassical black hole, including  all of the quantum effects of the bulk CFT state; while in the double holographic perspective the system can be studied by a higher dimensional black hole intersecting the brane\footnote{The double holography has been used to study quantum black hole\cite{Emparan:2002px}, island formula\cite{Almheiri:2019hni,Rozali:2019day,Chen:2020uac,Geng:2020qvw} and holographic complexity\cite{Hernandez:2020nem}.}. See Figure \ref{fig:intor} for illustration. 

Specifically, \cite{Emparan:2021hyr} explored the  volume complexity and action complexity of quantum corrected static BTZ using a  double holographic model, i.e., the static AdS$_4$ C-metric, and got some interesting results. We are going to generalize their studies to quantum corrected rotating BTZ black hole.  In this case, to study its holographic  complexity,  we need to use the rotating AdS$_4$ C-metric which is  of the form  \cite{Emparan:2020znc}
\begin{equation}
    \begin{aligned} \label{rot-qubtz}
       ds^2=\frac{\ell^2}{(\ell+xr)^2}\Big[&-\frac{H(r)}{\Sigma(x,r)}(dt+ax^2d\phi)^2+\frac{\Sigma(x,r)}{H(r)}dr^2\\
    &+r^2\Big(\frac{\Sigma(x,r)}{G(x)}dx^2+\frac{G(x)}{\Sigma(x,r)}(d\phi-\frac{a}{r^2}dt)^2\Big)   \Big],
    \end{aligned}
\end{equation}
where
\begin{equation}
    \begin{aligned}
       &H(r)=\frac{r^2}{\ell_3^2}+\kappa-\frac{\mu\ell}{r}+\frac{a^2}{r^2},\\
       &G(x)=1-\kappa x^2-\mu x^3+\frac{a^2}{\ell_3^2}x^4,\\
       &\Sigma(x,r)=1+\frac{a^2x^2}{r^2}.
    \end{aligned}
    \end{equation}
    The parameter $a$ characterizes the angular velocity of the black hole and we set it to be non-negative.
    This metric describes an Einstein-AdS$_4$ metric with AdS radius
    \begin{equation}\label{l-l3-l4}
        \ell_4=\Big(\frac{1}{\ell^2}+\frac{1}{\ell_3^2}\Big)^{-1/2},
    \end{equation}
    where $\ell_3$ is the constant curvature radius of AdS$_3$. The meaning of the parameter $\mu$ is a bit subtle and will be explained later. 

For the special case with $\mu=0$, the induced metric on the section $x=0$ is given by
\begin{equation}
ds^2\Big|_{x=0}=-\Big(\frac{r^2}{\ell_3^2}+\kappa+\frac{a^2}{r^2}\Big)dt^2+\frac{dr^2}{\frac{r^2}{\ell_3^2}+\kappa+\frac{a^2}{r^2}}+r^2\Big(\phi-\frac{a}{r^2}dt\Big)^2,
\end{equation}
 which is  a rotating BTZ black hole for $\kappa=-1$ or a conical defect in AdS$_3$  for $\kappa=1$. We are mainly interested in  the complexity of  quantum corrected rotating black hole, but also discuss the complexity of quantum corrected conical defect in AdS$_3$ for comparison. 
 The thermodynamic quantities of the black hole rely on the periodicity of $\phi$ which cannot naively be set to $2\pi$ but is determined by regularity condition. We will come to the details later.  
    
 We place the brane at $x=0$ and only keep the portion $x_1\geq x\geq0$. Then we glue this partial spacetime with its identical mirror through the brane via the Israel's junction conditions.  The junction condition requires that the metric is continuous at the gluing surface, while the extrinsic curvature could be discontinuous. This follows from the fact that on this surface there is a brane with tension $\tau$, and as a result the 4D Einstein equation has a source localized on the brane. 
    The extrinsic
curvature of the brane is given by
\begin{equation}
    K_{ij}=\frac{1}{2}\frac{\partial g}{\partial n}\Big|_{x=0}={-}\frac{1}{\ell}\tilde g_{ij},
\end{equation}
where $\partial_n=-r\partial_x$ and $\tilde g_{ij}$ is the induced metric on the brane at $x=0$. This extrinsic curvature satisfies the equation
\begin{equation}
    K_{ij}-\tilde g_{ij}K=4\pi G_4\tau\tilde g_{ij},
\end{equation}
which comes from the tension of brane  and the Einstein equations.   Therefore, the tension of the brane is  determined to be
    \begin{equation}
       \tau=\frac{1}{2\pi G_4\ell}.
    \end{equation}

For the more interesting case when we turn on a positive $\mu>0$, the induced geometry on the brane is no longer a locally AdS$_3$ but given by
\begin{equation}    \label{ads3-mu}
    ds^2=-\Big(\frac{r^2}{\ell_3^2}+ \kappa-\frac{\mu\ell}{r}+\frac{a^2}{r^2}\Big)dt^2+\frac{dr^2}{\frac{r^2}{\ell_3^2}+ \kappa-\frac{\mu\ell}{r}+\frac{a^2}{r^2}}+r^2\Big(\phi-\frac{a}{r^2}dt\Big)^2.
\end{equation}    
This metric is a solution of  Einstein-AdS$_3$ gravity  with higher curvature terms, coupled to  the holographic CFT$_3$ which is dual to the four-dimensional bulk. The effective action for this theory can be derived from the four-dimensional classical gravitational action with a brane, and is given by 
\begin{equation}\label{effact}
    I=\frac{1}{16\pi G_3}\int d^3x\sqrt{-h}\Big(R+\frac{2}{\ell_3^2}+\mathcal O(\ell^2)\Big)+I_{CFT},
\end{equation}    
 where the Netwon's constant in 3D effective theory is $G_3={G_4}/{2\ell_4}$ and $\mathcal O(\ell^2)$ terms represent the higher-curvature corrections. We will work in the regime where the Einstein-AdS terms dominate so that the metric \eqref{ads3-mu} can be viewed as a quantum corrected rotating BTZ black hole. This is true as long as $\ell\ll\ell_3$, which we assume hereafter. Moreover, as we have reviewed, the quantum matter is a holographic CFT$_3$ which is dual to AdS$_4$ with radius $\ell_4$ so that roughly the central charge $c_3$ satisfies
 \begin{equation}
 c_3\sim\frac{\ell_4^2}{G_4}\sim\frac{\ell}{G_3},
    \end{equation}
 where we have used $\ell_4\sim \ell$ when $\ell\ll\ell_3$ according to \eqref{l-l3-l4}.   Since $c_3$ counts the number of degrees of freedom in CFT$_3$, the action of CFT is of order $c_3$ while the typical value of gravitational action is  $\ell_3/G_3$. This allows us to define a dimensionless effective coupling as
 \begin{equation}
 g_{\mathrm{eff}}\sim \frac{I_{CFT}}{I_{grav}}\sim\frac{c_3G_3}{\ell_3}\sim\frac{\ell}{\ell_3},
\end{equation}   
which  is much smaller than $1$ when $\ell\ll\ell_3$. This confirms our expectation that in such case, the backreaction from the CFT$_3$ to AdS$_3$ is very small and can be viewed as a subleading correction. It is important to keep in mind that the leading CFT effect is of order $g_{\mathrm{eff}}\propto \ell$, while the higher-curvature corrections are of order $\mathcal O(\ell^2)$. 

When restricted to the case that the value of $x$ is positive, the domain of $r$ can be expressed as a union of two separated intervals. The first interval ranges from $-\infty$ to $-\ell/x$ which is the true asymptotic boundary of the AdS$_4$ geometry, while the second interval ranges from 0 to $+\infty$. In the bulk, the surface $r=\pm\infty$ has an AdS$_3$ geometry, and it can be accessed either from  the side where $r$ is positive and contains the brane, or from the side where $r$ is negative and extends to the boundary at the point where $rx=-\ell$. For the $\kappa=-1$ rotating black hole case, its Penrose diagram is shown in Figure \ref{fig:volumecom}.


The global properties and thermodynamics of the rotating quBTZ solution were extensively studied in \cite{Emparan:2020znc}. Here we provide a brief summary of the results, which will be useful for our study. To start with, we need to have both an outer and an inner horizon, $r_+$ and $r_-$, with $r_+>r_-$, in the rotating quBTZ solution. This requires an upper bound $a_{ext}$ on the parameter $a$,
\begin{equation}
a\leq a_{ext}=\frac{\ell_3}{2}+\mathcal{O}(\ell).
\end{equation}
We will also assume that we are working within a range of parameters where there is at least one positive root of $G(x)$. The smallest of these roots is referred to as $x_1$, in terms of which  $\mu$ can be expressed as
\begin{equation}
\mu=\frac{1-\kappa x^2_1+\tilde a^2}{x_1^3},\hspace{3ex}\mbox{with}~\tilde a=\frac{ax_1^2}{\ell_3}.
\end{equation}
We notice that $x=x_1$ is the fixed-point locus of the Killing vector
\begin{equation}\label{kill}
\hat{\xi}=\frac{\partial}{\partial\phi}-\tilde a\ell_3\frac{\partial}{\partial t_3}.
\end{equation}
To avoid a conical singularity at $x=x_1$, we must impose that $\phi\sim\phi+2\pi\Delta$, where
\begin{equation}
\Delta=\frac{2}{|G'(x_1)|}=\frac{2x_1}{3-\kappa x_1^2-\tilde a^2}.
\end{equation}
Additionally, the identification along the orbits of $\hat{\xi}$ must be made on the surfaces of constant $t+\tilde a\ell_3\phi$. As a result, the translation $\phi\to\phi+2\pi\Delta$ along an orbit of $\hat{\xi}$ does not return to the same point in spacetime but instead to another one at a different $t$. Specifically, we have the identification $(t,\phi)\sim({t}-2\pi\Delta\tilde a\ell_3,\phi+2\pi\Delta)$. With these considerations in mind, we can then transform to canonical coordinates $\bar{t}$ and $\bar{\phi}$ using the following coordinate transformation,
\begin{equation} \label{redf}
t=\Delta(\bar t-\tilde a\ell_3\bar\phi),\quad \phi=\Delta(\bar\phi-\frac{\tilde a}{\ell_3}\bar t).
\end{equation}
The above identification is given in terms of these new canonical coordinates by
\begin{equation}
(\bar t,\bar\phi)\sim(\bar t,\bar\phi+2\pi).
\end{equation}
The Killing vectors could be rewritten as
\begin{equation}
    \begin{aligned}
        &\frac{\partial}{\partial t}=\frac{1}{\Delta(1-\tilde a^2)}\Big(\frac{\partial}{\partial \bar t}+\frac{\tilde a}{\ell_3}\frac{\partial}{\partial \bar\phi}\Big),\\
          &\frac{\partial}{\partial \phi}=\frac{1}{\Delta(1-\tilde a^2)}\Big(\frac{\partial}{\partial \bar \phi}+\tilde a\ell_3\frac{\partial}{\partial \bar t}\Big).
    \end{aligned}
\end{equation}
As in \cite{Emparan:2020znc}, we require $\tilde a<1$ to avoid the existence of naked closed-timelike curve.  To conclude, the allowed range of parameters we will work in is
\begin{equation}
    a\leq a_{ext},\quad \tilde a=\frac{ax_1^2}{\ell_3}<1.
\end{equation}

After the coordinate transformation, the line element in the new coordinates $(\bar t,r,x,\bar\phi)$ becomes 
\begin{equation}\label{metric-can}
    \begin{aligned}
        ds^2&=\frac{\ell^2}{(\ell+xr)^2}\Big[-N^2d\bar t^2+\frac{\rho^2}{\Delta_r}dr^2+\frac{\rho^2}{\Delta_x}dx^2+\Phi^2(d\bar\phi-wd\bar t)^2\Big],
    \end{aligned}
\end{equation}
with     
\begin{equation}
    \begin{aligned}
    &\rho^2= r^2+a^2x^2,\quad\Delta_x=G(x),\quad\Delta_r=r^2H(r),\\
    & N^2=(1-\tilde a^2)^2\frac{\Delta^2\rho^2\Delta_r\Delta_x}{\bar\Sigma^2}, \\
    &\bar{\Sigma}^2=(r^2+a^2x_1^2)^2\Delta_x-a^2(x^2-x_1^2)^2\Delta_r,\quad\Phi^2=\frac{\Delta^2\bar\Sigma^2}{\rho^2},
    \end{aligned}
\end{equation}
\begin{equation}\label{w}
    w=-\frac{a}{\ell_{3}}\frac{(x^2-x_{1}^{2})(\ell_{3}^{2}-a^2 x^2 x_{1}^{2} )\Delta_{r}+(r^2+a^2 x_{1}^{2} ) (\ell_{3}^{2}+r^2 x_{1}^{2} )\Delta_{x} }{(r^2+a^2 x_{1}^{2})^2 \Delta_{x}-a^2 (x^2-x_{1}^{2} )^2 \Delta_{r}  }.
\end{equation}  
In this new coordinate system, the induced metric on the brane $x=0$ takes the standard form of asymptotically AdS$_3$, from which we can read off the mass $M$ and angular momentum $J$,
\begin{equation} \label{massang}
     M=\frac{\Delta^2}{8G_3}(1+\Tilde{a}^2+\frac{4\Tilde{a}^2}{x_1^2}),\quad
       J=\frac{\ell_3}{4G_3}\Tilde{a}\mu x_1\Delta^2.
\end{equation}
It then follows that\footnote{Note that there is a typo in \cite{Emparan:2020znc}.}
\begin{equation}
   8G_3\Big( M\pm\frac{J}{\ell_3}\Big)=\frac{4(1\pm\tilde a)^2(x_1^2\pm2\tilde a)}{(3+x_1^2-\tilde a^2)^2}.
\end{equation}
   The other thermodynamical quantities can be  determined as well,
    \begin{equation}\label{thermo}
    \Omega=\frac{a}{\ell_3}\frac{\ell_3^2+r_+^2x_1^2}{r_+^2+a^2x_1^2},~~~
       T=\frac{\Delta( 1-\Tilde{a}^2 )H'(r_+)}{4\pi(1+a^2x_1^2/r_+^2)},~~~
       S_{gen}=\frac{\pi}{G_4}\Delta\frac{\ell x_1(r_+^2+a^2x_1^2)}{\ell+r_+x_1}.
\end{equation}
In the limit of vanishing backreaction,  we have
\begin{equation}
    S_{gen}\Big|_{\ell=0}=S_{BTZ}=\frac{\pi\ell_3}{\sqrt{2G_3}}\Big(\sqrt{M+\frac{J}{\ell_3}}+\sqrt{M-\frac{J}{\ell_3}}\Big).
\end{equation}
In addition to the above quantities, one may associate a thermodynamic pressure with the cosmological constant in the framework of extended thermodynamics\cite{Kubiznak:2016qmn}, which in our case is given by
\be
P=\frac{3}{8\pi G_4\ell_4^2}.
\ee
Then the conjugate thermodynamic volume is defined as
\be\label{thermoV}
V\equiv\left(\frac{\partial M}{\partial P}\right)_{S,J}.
\ee
Therefore, the extended first law  of black hole thermodynamics is
\be\label{FL}
dM=TdS_{gen}+\Omega dJ+VdP.
\ee
 Using the first law \eqref{FL}, we get
\be
V=\left(\frac{\partial M}{\partial P}\right)_{a,x_1}-T\left(\frac{\partial S_{gen}}{\partial P}\right)_{a,x_1}-\Omega \left(\frac{\partial J}{\partial P}\right)_{a,x_1},
\ee
and find that the Smarr relation holds to the leading order
\be
M-(TS_{gen}+\Omega J)+2PV=O(\ell).
\ee

In the following, we will compute holographic complexity at a fixed time $\bar t$ using the CV and CA proposal respectively, and study its late-time behavior. In particular, we will check if the relation 
\be
\left.\frac{dC_{V,A}}{d\bar{t}}\right|_{\bar{t}\gg\b}\stackrel{?}{\sim} TS_{gen}.
\ee
holds universally or not.

 \section{Volume complexity}
 
\begin{figure}[h]
\centering
\includegraphics[scale=0.8]{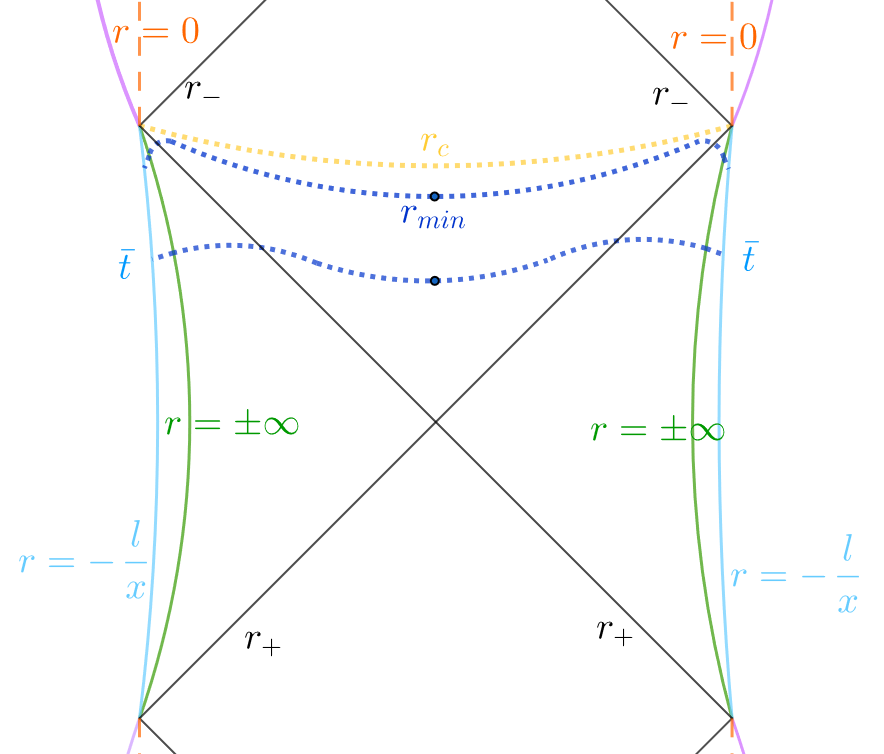}
\caption{Penrose diagram with maximal time slice for a constant $(x,\bar{\phi})$ section of the rotating $AdS_{4}$ C-metric geometry. The C-metric continues beyond $r=\pm \infty$ into negative $r$ until it comes to the asymptotic boundary $r=-\frac{\ell}{x}$. Two dotted blue lines represent maximal wormholes connecting two asymptotic boundaries at different boundary time $\bar{t}_{L}=\bar{t}_{R}=\bar{t}$ (depicted at symmetric time in the figure). The minimal radius $r_{min}$ is defined through $\dot{\bar{t}}^{-1}=0$. For the late-time regime, the maximal time slice approaches constant $r_{c}$ surface (yellow), which is determined by $\eqref{rc}$.}
\label{fig:volumecom}
\end{figure}

In this section, we compute the volume complexity of the rotating quBTZ black hole at constant time $\bar t$, which is given by
\be \label{cv4}
C_V(\bar t)=\frac{\text{Vol}(\Sigma_{\bar t})}{4\pi G_4\ell_3}.
\ee
where we have chosen $L=4\pi\ell_3$ in \eqref{def:CV}.
The surface $\Sigma_{\bar t}$ is anchored to the left and right asymptotic boundaries at $\bar t_L=\bar t_R=\bar t$,  and should maximizes the volume functional, see Figure $\ref{fig:volumecom}$. Actually, we can also consider the  holographic complexity at a fixed time $t$ in the original coordinates. Then we find that the late-time derivative of CV differs from the one at constant $\bar t$ by a factor. However,  the WdW patch of constant-time slice in the original coordinates is not well defined. We address these points in the appendix B.

The above volume complexity includes the quantum effect of CFT$_3$. It would be illuminating to  expand \eqref{cv4}  in terms of three-dimensional semiclassical quantities
\be \label{phyexp}
C_V(|\Psi \rangle)=\frac{\mbox{Vol}(\Sigma)}{G_3 L}+\frac{ \delta \mbox{Vol}(\Sigma) + \mathcal{V}(\Sigma) }{G_3  L}+C_{V}^{bulk}(|\phi\rangle)+...\quad .
\ee 
The first term is the classical part, and the second and third terms are the leading quantum corrections from CFT matter.  The state $|\Psi \rangle$ represents the UV complete quantum state whose complexity we try to compute, while the state $|\phi\rangle$ is the coarse-grained semi-classical bulk quantum state defined on the curved geometry. The effective action of three-dimensional theory includes higher curvature terms, whose leading order effects are revealed in $\mathcal{V}(\Sigma)$ term and are of order $\mathcal{O}(\ell^2)$. The bulk quantum fields $|\phi\rangle$ can backreact on the geometry causing a change in the extremal hypersurface volume $\delta \mbox{Vol}(\Sigma)$, giving rise to the correction of the leading order $\mathcal{O}(\ell)$. Also the semiclassical state itself would contribute to the complexity $C_V^{bulk}$ in a proper way, but its contribution is subleading. At last, the dots generally include higher-curvature terms in the effective action and higher order terms in $g_{\text{eff}}$.    

\subsection{Extremal volume}
We proceed the calculation in an analogous way to \cite{Emparan:2021hyr} and introduce a new coordinate
     \begin{equation}
       z=xr,
   \end{equation}
   and use $x^\mu=(\bar t,r,z,\bar\phi)$ as the coordinates of spacetime. Using the translational symmetry along $\phi$ direction, the extremal surface can be parametrized as $\bar t=\bar t(z,r)$. In these coordinates, the induced metric of $\Sigma_{\bar t}$ is given by
   \begin{equation} \label{angudef}
       \begin{aligned}
           ds^2\big|_{\Sigma_{\bar t}}=&\frac{\ell^2}{(\ell+z)^2}\Big[\Big(-N^2\Dot{\bar t}^2+\frac{\rho^2}{\Delta_r}+\frac{z^2\rho^2}{r^4\Delta_x}\Big)dr^2+\Big(-N^2(\bar {t}')^2+\frac{\rho^2}{r^2\Delta_x}\Big)dz^2\\
           &-2\Big( N^2\Dot{\bar t} \, \bar t'+\frac{z\rho^2}{\Delta_xr^3}\Big)drdz+\Phi^2 {(d\phi-wd\bar t)^2} \Big]
       \end{aligned}
   \end{equation}
   with $\dot{\bar t}=\partial_r\bar t,\; \bar t'=\partial_z\bar t$ and $d\bar t=\dot{\bar t}dr+\bar t'dz$.  
The volume complexity of the system  is then obtained by finding the extremal value of the functional
    \begin{equation}\label{CV}
        C_V(\bar t)=\text{ext}\left\{\frac{\ell}{ {2} G_4\ell_3}\int drdz\frac{\ell^2}{(\ell+z)^3}\sqrt{\mathcal{V}}\right\}
    \end{equation}
    with
 \begin{equation} \label{limit}
 \begin{aligned}
     \mathcal{V} & =\frac{ \Phi^2\rho^2(r^2\rho^{{2} }-N^2(\Delta_r(r\dot{\bar t}+z\bar t')^2+r^4\Delta_x\bar t'^2))}{r^4\Delta_r\Delta_x}\\
     &= -\frac{(r^4+a^2 z^2) \Delta^{2} }{G H r^{8} \rho^{2} } \big(-G (r^3 + a^2 r x_{1}^{2} )^2+a^{2} H (z^2-r^2 x_{1}^{2})^2\\
     &\quad + G^2 H r^6 {\bar t}'^{2} \Delta^{2}(1-\Tilde{a}^{2})^2 + G H^2 r^4  \Delta^{2} (r \dot{\bar t}+z {\bar t}' )^2 (1-\Tilde{a}^{2})^2 \big)
 \end{aligned}   
 \end{equation}

There are two remarkable points on the function $\eqref{limit}$. The first point is  its discontinuous behavior in the zero angular-momentum limit. In this limit there is 
\begin{equation}  \label{vollt}
\mathcal{V}|_{a=0}= r^2  \Delta^{2} \( H^{-1} -H \( \dot{\bar t}+ \frac{z {\bar t}'}{r} \)^2 \Delta^{2}-G {\bar t}'^2 \Delta^{2} \),
\end{equation}
which does not reduce to the one of the static quBTZ \cite{Emparan:2021hyr}. The discontinuity originates from   the redefinition of time and angular coordinates in $\eqref{redf}$ as well as the fixing of the time $\bar{t}$ coordinate we choose to evaluate the holographic volume complexity. It is not hard to see that in $\eqref{vollt}$ if we redefine $\Delta t$ as the new time coordinate, we would recover the result $(3.5)$ in \cite{Emparan:2021hyr}, which is exactly the manifestation of redefinition in $\eqref{redf}$.  The second point is more technical and subtle. The large $z$ behavior of $\eqref{limit}$ is
\begin{equation}
   \mathcal{V}|_{z \rightarrow \infty} \sim z^{8} 
\end{equation}
which would  ruin the key formula $\eqref{expansion}$. However after checking it carefully, we find that its effect is of order $\ell^2$ and can be ignored safely in our investigation, as shown below. 

Recall that we are actually considering the two-sided symmetrical spacetime. Though the AdS C-metric could be defined by doing analytically continuation beyond $r=0$ into negative values of the radial coordinate,  it is more convenient to  consider the $r>0$ part of the spacetime and symmetrically double it. 
The boundary conditions in the extremization problem about $\eqref{CV}$ are of Dirichlet type at $z=- \frac{\ell}{x}$ for two-sided asymptotic boundaries and of Neumann type at the intersection of the brane and the extremal volume slice $\Sigma_{\bar t}$ due to the tensional character of the brane. The Neumann boundary condition can be seen from the extremality condition not away from the brane, but on the brane. The surface element $dS_{a}$ should be tangent to $\Sigma_{\bar t}$, or put it another way,  the normal vector  $N_{\Sigma_{\bar t}} \propto (d\bar t-\Dot{\bar t}dr-\bar t'dz)$ should lie on the brane at $z = 0$. Mathematically, we need 
\be \label{bd1}
   g_{\mu \nu} N_{brane}^{\mu} N_{\Sigma_{\bar t}}^{\nu}=0, \quad N_{brane} \propto dz \quad \longmapsto \quad \bar t'(r,0)=0.
\ee

As we assume that the quantum correction is small, $\ell \ll \ell_3$, we may first consider the small $\ell$ limit of $C_V$.  Keep in mind that $G_4\sim2\ell G_3$, so the prefactor of \eqref{CV} is of order 1. For  the first factor of the integrand in \eqref{CV},  we claim that in the small $\ell$ limit, the following identity holds
 \begin{equation}\label{expansion}
      \frac{\ell^2}{(\ell+z)^3}=\delta(z)-\ell\frac{d}{dz}\delta(z)+O(\ell^2)
    \end{equation}
 in the sense of distribution. This is indeed the equality used in \cite{Emparan:2021hyr} but there is a subtle difference in our case. In \cite{Emparan:2021hyr}, \eqref{expansion} is proved using a test function which grows no faster than $z^2$ as $z\to\infty$. As we mentioned before, the test function for the rotating case is $\sqrt{\mathcal{V}}\sim z^4$. Nevertheless, \eqref{expansion} still holds up to the linearized order of $\ell$ so that we can use it for our computation.
 To prove this fact, we note that when
    \be f(z)|_{z \to \infty} \sim z^{4},  \label{largezf}
    \ee
    there is 
    \be\ba
     I&=2\int_{0}^{\infty} dr \int_{0}^{x_{1}} dx \frac{\ell^2}{(\ell+rx)^3} f(rx)=2\int_{0}^{\infty} dr \int_{0}^{x_{1}r} dz \frac{\ell^2}{(\ell+z)^3} f(z)\\
     &\sim -\frac{\ell^2 f(z)}{(\ell+z)^2}\bigg\vert_{0}^{x_{1} r }-\frac{\ell^{2} f'(z) }{\ell+z}\bigg\vert_{0}^{x_{1} r}+\int_{0}^{x_{1}r} \frac{\ell^2 f''(z)dz}{\ell+z}. 
     \ea
     \ee
    In the last step we have used the integration by parts and ignored the $dr$ part of the integration. We then find
   \be I\sim f(0)+\ell f'(0)+\(\ell^2 (x_{1} r)^2-12 \ell^3 r x_1+12 \ell^4  \log{(\ell+r x_1)}\), \ee
    where we have used the large-$z$ behavior of $f(z)$ \eqref{largezf}. We see that although there are three divergent terms, one being linearly divergent, one being quadratically divergent and one being  logarithmically divergent,  all of them are beyond the order $\ell^2$,   whose effects will be ignored in this work.  Also it can be directly checked that the functional $\mathcal V$ satisfies
    \begin{equation}\label{property}
       \delta(z) \frac{d}{dz}\sqrt{\mathcal{V}}=0
    \end{equation}
    using the boundary condition \eqref{bd1} and $G'(0)=0$.
 This means that the subleading term in the expansion \eqref{expansion} does not contribute to the integration.
Therefore, the leading-order quantum correction to the integration \eqref{CV} totally comes from the integrand $\mathcal V$, and \eqref{CV} reduces to the integration on the brane $z=0$. 
Remember that we have only considered $r>0$ part of the integral so far, which has classical contribution because it contains the brane $x=0$ which can make $z=xr=0$. The $r<0$ part does not contribute classically since $z=xr\in(-\infty,-\ell)$ and the leading delta function in \eqref{expansion} excludes the contribution from negative value of $z$. 
Evaluating \eqref{CV} and expanding the result to the linearized order we get
    \begin{equation}
    \begin{aligned}\label{CV2}
        C_V(\bar t)&=\frac{\ell }{2G_4\ell_3} \text{ext} \left\{   \int dr\sqrt{\mathcal V}\big|_{z=0} \right \} 
        =\frac{\ell \Delta}{2G_4\ell_3}  \text{ext}\left\{   \int dr\sqrt{-F_1\Dot{\bar t}^2+F_2}  \right\} 
    \end{aligned}
    \end{equation}
    where
    \begin{equation} \label{f1f2}
    \begin{aligned}
          & F_1=(1-\tilde a^2)^2\Delta^{2}\Delta_r,\quad  \quad   \quad F_2=\frac{{(r^2+a^2 x_{1}^{2})^2-a^2 x_{1}^{4} \Delta_{r}}}{\Delta_r}.
    \end{aligned}
    \end{equation}
   By extremizing the functional, we find  one conserved quantity
    \begin{equation}
    E_t=\frac{F_1\Dot{\bar t}}{\sqrt{-F_1\Dot{\bar t}^2+F_2}}
    \end{equation}
    which implies
    \begin{equation}\label{conserve}
    \Dot{\bar{t}}= - \frac{E_t\sqrt{F_2}}{\sqrt{F_1(E_t^2+F_1)}}.
    \end{equation}

Due to the reflection symmetry in the radial coordinate, the extremal surface decreases from the left boundary to the interior and reaches its minimum $r_{min}$ at the turning point at time $\bar t=0$, after which the radial coordinate increases till to the right boundary. At the turning point, we have $\dot{\bar t}^{{-1}}=0$, so
    we can write $E_t$ in terms of the minimum radius $r_{min}$ of the extremal surface as
    \begin{equation}\label{Et}
        E_t=\sqrt{-F_1(r_{min})}.
    \end{equation}
    Integrating both sides of \eqref{conserve}, we obtain
    \begin{equation}\label{t-rmin}
    \bar t=-\int^{r_\infty}_{r_{min}}\frac{E_tdr}{\sqrt{F_1F_2^{-1}(E_t^2+F_1)}}.
    \end{equation}
   Substituting \eqref{conserve} into \eqref{CV2}, we obtain the volume complexity
     \begin{equation}\label{CV3}
     C_V=\frac{2\ell\Delta}{G_4\ell_3}\int^{r_\infty}_{r_{min}}dr\sqrt{\frac{F_1F_2}{E_t^2+F_1}},
    \end{equation}
 which depends on $r_{min}$. Considering the fact that Eq. \eqref{t-rmin} gives the relation between $\bar t$ and $r_{min}$, we see the time dependence of the volume complexity. However, as the integration in \eqref{t-rmin} and \eqref{CV3} cannot be performed analytically,  we do not have an explicit expression for the full time dependence. Nevertheless, we can study the early-time and late-time behavior of the complexity  at the linearized order of $\ell$.

Before we discuss the time-dependence of the complexity, we would like to discuss the quantum corrections in the complexity. We may formally expand the volume complexity as
    \begin{equation} \label{mathexp}
        C_V(  {\bar t} )=C_V^0(  {\bar t} )+\delta C_V(  {\bar t} )+\mathcal{O}(g_{\mathrm{eff}}^2)
    \end{equation}
 where $C_V^0(  {\bar t} )$ is the classical contribution while $\delta C_V(  {\bar t} )$ is the leading quantum correction.  
 Setting $\ell=0$ in \eqref{CV2} gives the classical part of the volume complexity
 \be\label{CV0}
 C_V^0=\frac{\Delta}{4G_3\ell_3}  \text{ext} \left\{   \int dr\sqrt{-F_1^{(0)}\dot{\bar t}^2+F_2^{(0)}}\right\}
 \ee
 where
    \begin{equation}\label{0order}
    \ba
 &F_1^{(0)}=(1-\tilde a^2)^2\Delta^2\Delta_r^{(0)},\quad F_2^{(0)}=\frac{F^{(0)}}{\Delta_r^{(0)}},\\
 &\Delta_r^{(0)}=\frac{r^4}{\ell_3^2}-r^2+a^2,\quad F^{(0)}=(1-\tilde a^2)r^4+a^2x_1^2 r^2 (2 +x_1^2).
 \ea
    \end{equation}
 This is precisely  the volume  complexity of constant $\bar t$ slice in the geometry of a BTZ black hole. To see this, we note that under the limit $\ell\to0$, the induced metric on the brane $x=0$ is given by
 \be
 \ba\label{ind-met0}
 \lim_{\ell\to0}ds^2\Big|_{x=0}=-\frac{r^2F_1^{(0)}}{F^{(0)}}d\bar t^2+\frac{r^2}{\Delta_r^{(0)}}dr^2+\frac{\Delta^2F^{(0)}}{r^2}(d\bar\phi-wd\bar t)^2.
 \ea
 \ee 
 Therefore, it is easy to see that  \eqref{CV0} computes the volume of the  extremal surface $\bar t=\bar t(r)$ in the BTZ geometry described by \eqref{ind-met0}. As a result, we have
 \be 
 C_V^0=C_V^{BTZ}(\bar t)
 \ee

As argued below \eqref{phyexp}, there are two contributions to the  leading quantum correction $\delta C_V$ in \eqref{mathexp}.  Firstly, the higher curvature correction to the three dimensional effective action is at least of order $\ell^2$ by virtue of \eqref{effact}, so the $\mathcal{V}(\Sigma)$ term in $\eqref{phyexp}$ would not appear in $\delta C_V$. Secondly,  the correction from  CFT$_3$ bulk fields can be further divided into two parts. One part is the UV divergent part $C_{UV}$ which is just given by the $r<0$ part of the integral \eqref{CV} and it is interpreted as the complexity of the CFT$_3$ bath. The other part would be the  complexity of the CFT$_3$ on the gravitating black hole geometry. However,  due to the property \eqref{property}, the leading-order correction to the complexity totally comes from the correction to the volume functional, which should be attribute to the backreaction of the CFT$_3$ to the geometry. This means that the proper complexity of this particular state of dCFT in our double holography model does not contribute at the linearized order, i.e. $ C_{V}^{bulk}(|\phi\rangle)=\mathcal{O}(g_{\text{eff}}^{2})$ in $\eqref{phyexp}$. As a result, we can write
 \be
 \delta C_V{(\bar{t})}=C_{UV}{(\bar{t})}+ {\frac{ \delta \text{Vol}(\Sigma_{\bar{t}}) }{8 \pi G_{3} l_{3}}} ,
 \ee
 where $\delta \mbox{Vol}(\Sigma_{\bar{t}})$ is the $\mathcal{O}(\ell)$ change in the volume of the extremal surface $\Sigma_{\bar{t}}$ on the brane by the semiclassical backreaction of CFT$_{3}$.  The UV-divergent part $C_{UV}$ gets contribution from  $r<0$ region which includes the asymptotic boundary, and is given by   
 \be
 C_{UV} {(\bar{t})} ={\lim_{\epsilon \to 0}}  \frac{\ell}{G_4\ell_3}\int_{-\infty}^{-\frac{\ell+\epsilon}{x_1}}dr\int^{x_1r}_{-\ell+\epsilon}dz\frac{\ell^2}{(\ell+z)^3}\sqrt{\mathcal V} {\bigg\vert_{\Sigma_{\bar t}}},
 \ee
 where we have introduced a bulk regulator near the boundary at $z=-\ell+\epsilon$, and  the additional factor of two compared to \eqref{CV} comes from two copies in double holography. Note that this term receives the contribution only outside the black hole and is time independent, whose general structure can be analyzed following \cite{Carmi:2016wjl} when $\epsilon \to 0$. Therefore it is not of interest to us. 
 In the following, we just focus on $\delta \mbox{Vol}(\Sigma_{\bar{t}}) $ and compute both the early- and late-time growth rates  of the volume complexity.

\subsection{Early-time behavior}
  At $\bar t=0$, we should have $E_{\bar t}=0$ by using \eqref{t-rmin}. This further implies the extremal surface reaches its minimal radius at the horizon $r_{min}=r_+$ according to \eqref{Et}. The complexity \eqref{CV3} now becomes
  \be
  C_V=\frac{2\ell\Delta}{G_4\ell_3}\int^{r_\infty}_{r_{min}}dr\sqrt{F_2}.
  \ee
  To compute the quantum correction, we can expand $F_2$ and $r_{min}$ in orders of $\ell$ as
  \be
  F_2=F_2^{(0)}+F_2^{(1)}+O(g_{\mathrm{eff}}^2),\quad r_{min}=r_{min}^{(0)}+r_{min}^{(1)}+O(g_{\mathrm{eff}}^2)
  \ee
  where $F_2^{(0)}$ is given by \eqref{0order} and 
\begin{equation}\label{rpapp}
\begin{aligned}
&F_{{2}}^{( {1})}=\frac{r(r^2+a^2x_1^2)^2\mu {\ell}}{(\Delta_r^{(0)})^2},\\
& r_{min}=   r_+=\sqrt{\frac{\ell_3^2+\ell_3\sqrt{\ell_3^2-4a^2}}{2}}+\frac{\ell_3\mu\ell}{2\sqrt{\ell_3^2-4a^2}}+O(g_{\mathrm{eff}}^2)=r_{min}^{(0)}+ r_{min}^{(1)}+O({\ell}^2).
 \end{aligned}
\end{equation}
Therefore, the contribution from the backreaction of CFT$_3$ to the volume complexity is given by
\begin{equation}
    \begin{aligned}
         C_{\delta V}&=\frac{\ell\Delta}{G_4\ell_3}\Big(\int^{r_\infty}_{r_{min}^{(0)}}dr\frac{F_2^{(1)}}{\sqrt{F_2^{(0)}}}-2\sqrt{F_2^{(0)}(r_{min}^{(0)})} r_{min}^{(1)}\Big)\\
        &=\frac{\ell\Delta}{G_4\ell_3}\Big[\int^{r_\infty}_{r_{min}^{(0)}}\frac{F_2^{(1)}}{\sqrt{F_2^{(0)}}}\Big(1{+}\frac{a^2\ell_3^2(2+x_1^2)-r^2(\ell_3^2+2a^2x_1^2)}{\ell_3\sqrt{\ell_3^2-4a^2}(r^2+a^2x_1^2)}\Big)dr-2\sqrt{\ell_3^2-a^2x_1^4}\ r_{min}^{(1)}\Big]
    \end{aligned}
\end{equation}
where in the second line, we have used 
\be
\sqrt{F_2^{(0)}(r_{min}^{(0)})}=\sqrt{F_2^{(0)}(r_\infty)}-\int^{r_\infty}_{r_{min}^{(0)}}\frac{d}{dr}\sqrt{F_2^{(0)}(r)}.
\ee
The factor in the parenthesis of the integrand is a decreasing function of $r$ and reaches its maximum at $r_{min}^0=0$. Therefore, the integrand is non-positive which implies $ C_{\delta V}<0$ , i.e.  the overall backreaction correction is negative at very early time.

\subsection{Late-time behavior}
    The derivative of the complexity with respect to the anchoring time $\bar t$ is given by the Hamilton-Jacobi equation for $C_V$, 
    \begin{equation}
        \frac{dC_V}{d\bar t}=\frac{\ell\Delta}{2G_4\ell_3}\Big(\frac{\mathcal{L}_V}{\partial \Dot{\bar{t}}}\Big|_{r=r_\infty^L}+\frac{\mathcal{L}_V}{\partial \Dot{\bar{t}}}\Big|_{r=r_{\infty}^R}\Big)=\frac{\ell\Delta}{G_4\ell_3}|E_{\bar t}|
    \end{equation}
    where $\mathcal L_V:=\sqrt{\mathcal V}$.
    At late time, $r_{min}$ approach to the critical value $r_c$ which extremizes $\sqrt{-F_1(r_{min})}$, i.e.
    \begin{equation} \label{rc}
        \frac{d}{dr}(\sqrt{-F_1})\Big|_{r_c}=0.
    \end{equation}
  Up to the linearized order of $\ell$, such critical radius is solved to be
        \begin{equation}
        r_c=\frac{\ell_3}{\sqrt{2}}+\frac{\ell\mu}{4}+O(\ell^2).
    \end{equation}
    Then the growth rate of the complexity is given by
\begin{equation}
\begin{aligned}\label{CV-final}
     \frac{dC_V}{d\bar{t}}&=\frac{\ell\Delta^2(1-\tilde a^2)}{G_4\ell_3}r_c\sqrt{-H(r_c)}
\\&=\frac{\Delta^2(1-\tilde a^2)}{ 4 G_3\ell_3}\sqrt{\ell_3^2-4a^2}\Big(1+\frac{\sqrt{2}\ell_3\mu\ell}{\ell_3^2-4a^2}\Big), 
\end{aligned}
\end{equation}
where we have used $G_4=2G_3\ell_4$.
Setting $a=0$, we can get the same rate of volume complexity of non-rotating quBTZ black hole as in \cite{Emparan:2021hyr}.

In \cite{AlBalushi:2020heq,AlBalushi:2020rqe}, the growth rate of the holographic complexity of odd-dimensional Myers-Perry black hole is shown to be the difference in internal energies  between the inner and outer horizons. In the large blackhole limit where the horizon radius $r_+$ is much larger than the AdS radius, it was found
\be
\frac{dC_{V/A}}{dt}\propto P\Delta V
\ee
where $\Delta V=V-V_-$ with $V_-$ being defined on the inner horizon in an analogous way to the thermodynamic volume $V$ as in \eqref{thermoV}.
Since we are considering the small $\ell$ limit,  it is also in the large blackhole limit $r_+\gg\ell_4$. Actually, it is direct to check that in our situation, the following equations
\be
TS_{gen}-T_-S_-\propto P\Delta V,\quad TS_{gen}=-T_-S_-
\ee
hold to the classical order, where $T_-$ and $S_-$ are the temperature and entropy defined on the inner horizon. At the quantum order,  both $P\Delta V$ and $TS_{gen}-T_-S_-$ are complicated and differ from \eqref{CV-final}. Consequently, it is expected that 
\be
\frac{dC_{V/A}}{dt}\propto TS_{gen}
\ee
at the leading order. 

To compare our result of volume complexity with the above thermodynamic quantity, 
 we shall only use  $TS_{gen}$ for simplicity and expand it to linear order by using \eqref{thermo} and \eqref{rpapp}
\begin{equation}\label{TS}
    TS_{gen}=\frac{\Delta^2\sqrt{\ell_3^2-4a^2}(1-\tilde a^{{2}})}{4G_3\ell_3}(1+\frac{\ell}{\ell_3} F_{gen}(x_1,\tilde a)).
\end{equation}
The function $F_{gen}(x_1,\tilde a)$ is of such a complicated expression that we would not show it explicitly. We may define a function $F_V$ in an analogous way as \eqref{TS} to rewrite
 \begin{equation}
\begin{aligned}\label{CV:dt}
     \frac{dC_V}{d\bar t}&=\frac{\Delta^2\sqrt{\ell_3^2-4a^2}(1-\tilde a^2)}{4G_3\ell_3}(1+\frac{\ell}{\ell_3}F_V(x_1,\tilde a))
     \end{aligned}
\end{equation}
Similar to the non-rotating quBTZ case, although  $F_V$ and $F_{gen}$ differ, the ratio between them is of order one.


The late-time behavior of volume complexity of rotating quBTZ has a richer structure than the one of static case, as there are more parameters in the metric, which may change the root structure of the blackening factor $H(r)$.  This becomes more apparent if we consider the case that $\kappa=1$, which corresponds to quantum-corrected conical defect. Recall that in the static case with $\kappa=1$, $H(r)$ always has a real positive root of order $\ell$, implying a highly quantum dressed horizon.
In the rotating quBTZ case, there are two free parameters $\ell$ and $a$ in the blackening factor $H(r)$ of C-metric \eqref{rot-qubtz}, where $\ell$ characterizing the quantum backreaction of CFT$_3$ and $a$ denoting the angular momentum of black hole\footnote{Note here we explore the complexity at a fixed parameter $a$ rather than a fixed angular momentum $J$.}. Therefore, for the rotating case with $\kappa=1$, whether $H(r)$ has positive roots depends on the range of $a$. More precisely, there is an upper bound on $a$ beyond which $H(r)$ has no real root indicating no horizon at all. The upper bound can be obtained by noting that the existence of horizons is equivalent to the following condition
\be \label{arange} \text{Min} \( H(r)|_{r>0}\) <0, \ee
Solving above equation to the leading order in $\ell$, we get
\be \label{bound:a}
a_0\equiv \frac{a}{\ell}<\frac{\mu}{2}\sim \frac{1-x_{1}^2}{2 x_{1}^{3}}, \ee
whose range depends on the parameter $x_{1}$.  The upper bound is not a constant as we change $x_{1}$. 
When \eqref{bound:a} is satisfied, $H(r)=0$ has two positive roots, just like the case of usual rotating black hole. The growth rate of the extremal volume is again computed by the first line of  \eqref{CV-final} with $\kappa=1$, and the result is
\be\label{CV:defect}
\frac{dC_V}{d\bar t}=\frac{ \Delta^2  \ell\sqrt{\mu^2-4a_0^2}}{4G_3\ell_3}+O(\ell^2)
\ee
In the limit $a_0\to0$, we recover the result for non-rotating quantum BTZ. Since the growth rate is of order $\ell$, it has no classical counterpart.  This is expected since the horizon radius is of order $\ell$ so the black hole is quantum in origin. When $a\gg \ell$, it is easy to see  that $H(r)$ is always positive for all $r>0$  and so is $F_1(r)$ in \eqref{f1f2}. As a result, there will be no turning point for the extremal surface with $\dot{\bar t}^{-1}=0$, otherwise the  conserved   energy in \eqref{Et} would be pure imaginary. In other words, in this case, the extremal condition is not compatible with the boundary condition, so there is  no extremal surface. Nevertheless, it is easy to find the surface which maximizes the volume functional that is simply $\bar t=$const. This implies that $C_V$ is constant in time.

\begin{figure}
    \centering
    \includegraphics[scale=0.6]{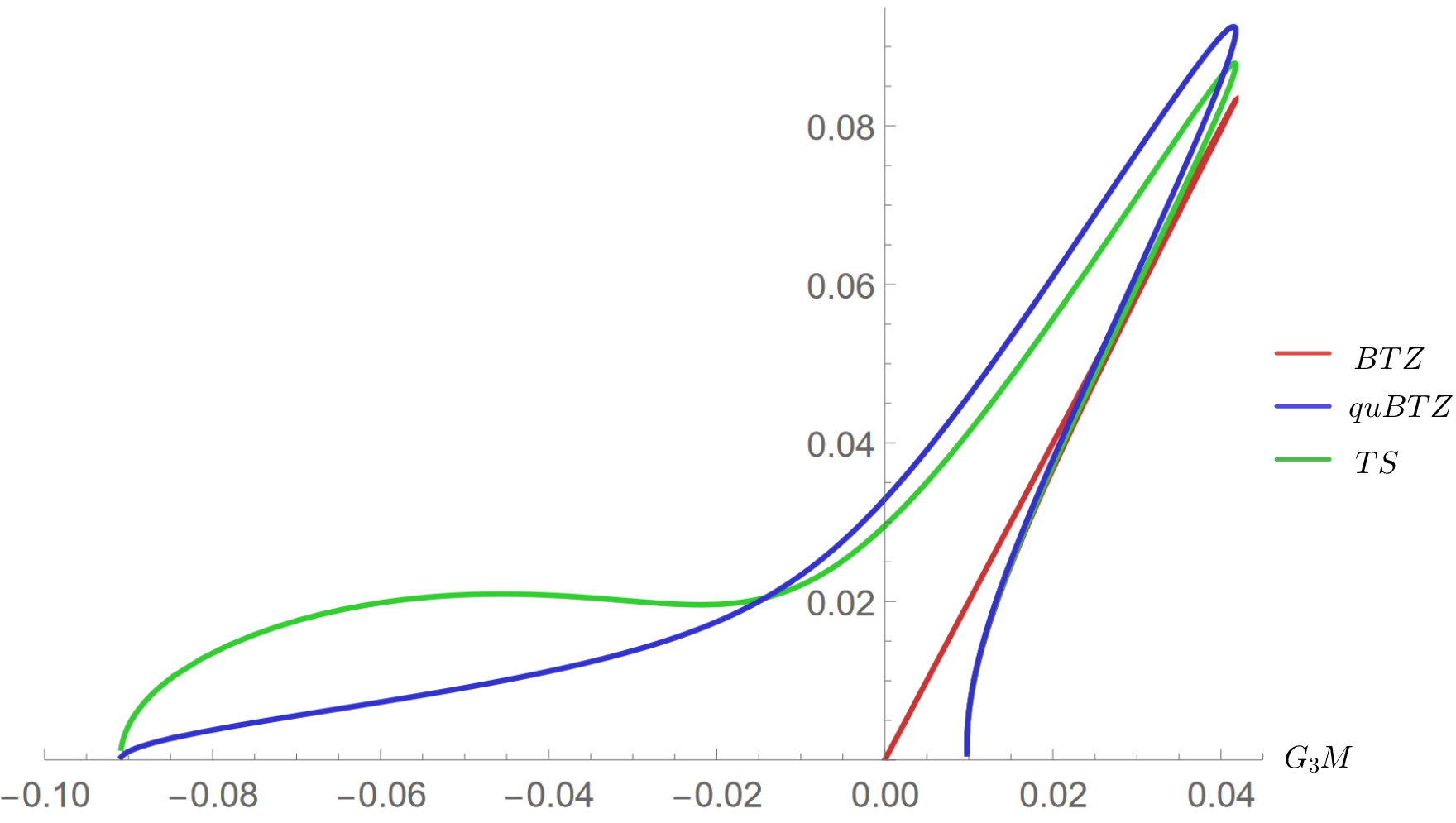} 
    \caption{The late-time behavior of volume complexity of rotating quBTZ (blue line), the one of classical rotating BTZ black hole (red line),  and the thermodynamic quantity $TS_{gen}$ (green line). For the rotating quBTZ,  we set $\ell=0.1,a=0.01$ and $\ell_3=1$. }
    \label{fig:volallval}
\end{figure}

In Figure \ref{fig:volallval}, we compare the late-time behavior of volume complexity of rotating qu-BTZ, the one of classical rotating BTZ black hole,  and the thermodynamic quantity $TS_{gen}$. In the figure, we include the study of the volume complexity of rotating quantum conical defects with $M<0$, which is described by the AdS$_4$ metric \eqref{rot-qubtz} with $\kappa=1$. Similar to the static case, the rotating quBTZ ($M>0$) reproduces the behavior of $TS_{gen}$ up to $\mathcal{O}(1)$ coefficients in the leading quantum correction. Another remarkable point is that for the conical case, its late-time behavior of volume complexity is not simply proportional to the quantity $TS_{gen}$.

\begin{figure}[htbp]
\centering
\includegraphics[width=0.45\textwidth]{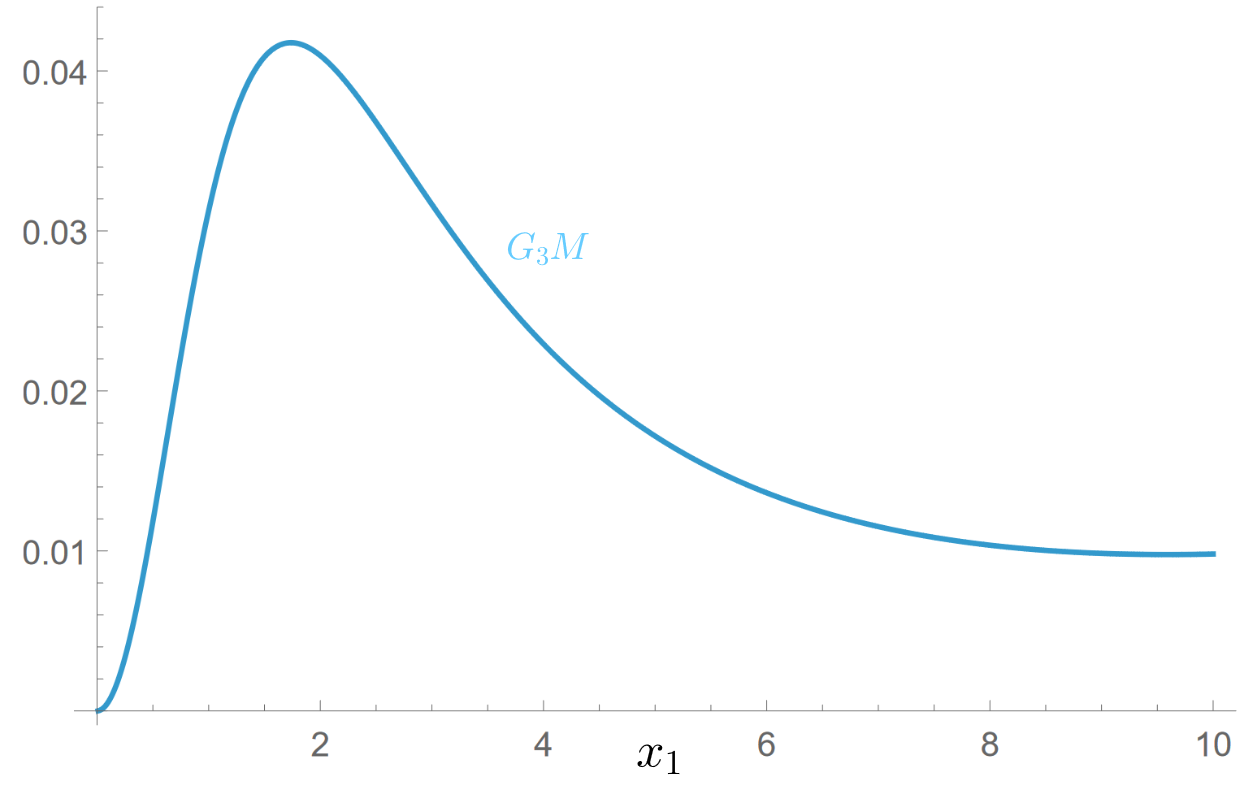}
\qquad
\includegraphics[width=0.45\textwidth]{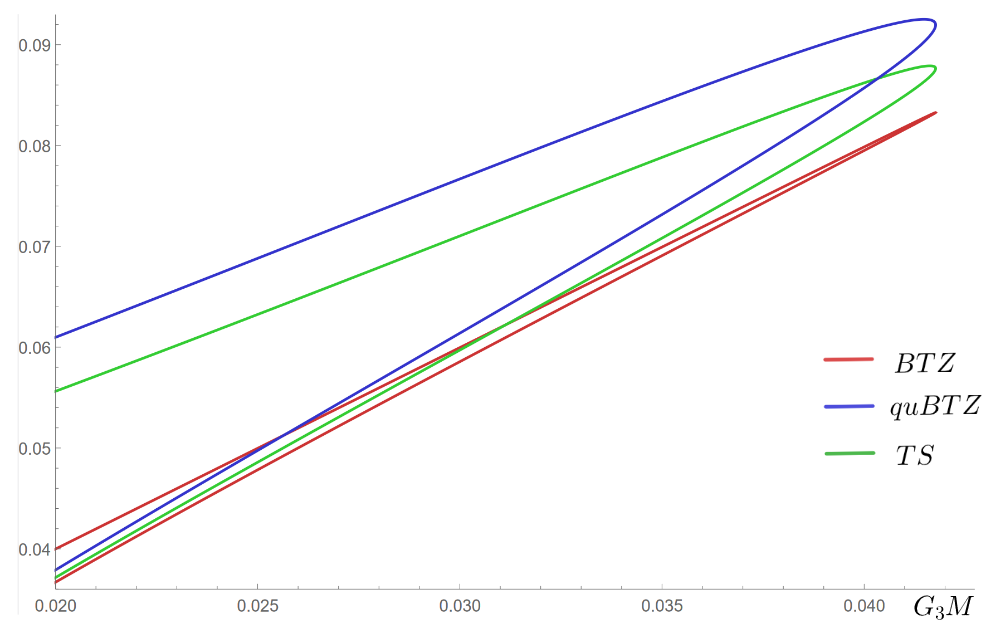}
\caption{The late-time behaviors of rotating quBTZ and rotating BTZ, as well as $TS_{gen}$,  for $\ell=0.1,a=0.01$ and $\ell_3=1$. The left figure shows the  dependence of $M$ on $x_1$. We can see that for some range of $M$, there are two branches of $x_1$ and for the second branch, $M$ does not go back to zero. }
\label{fig:volmass}
\end{figure}

\begin{figure}[htbp]
\centering
\includegraphics[width=0.4\textwidth]{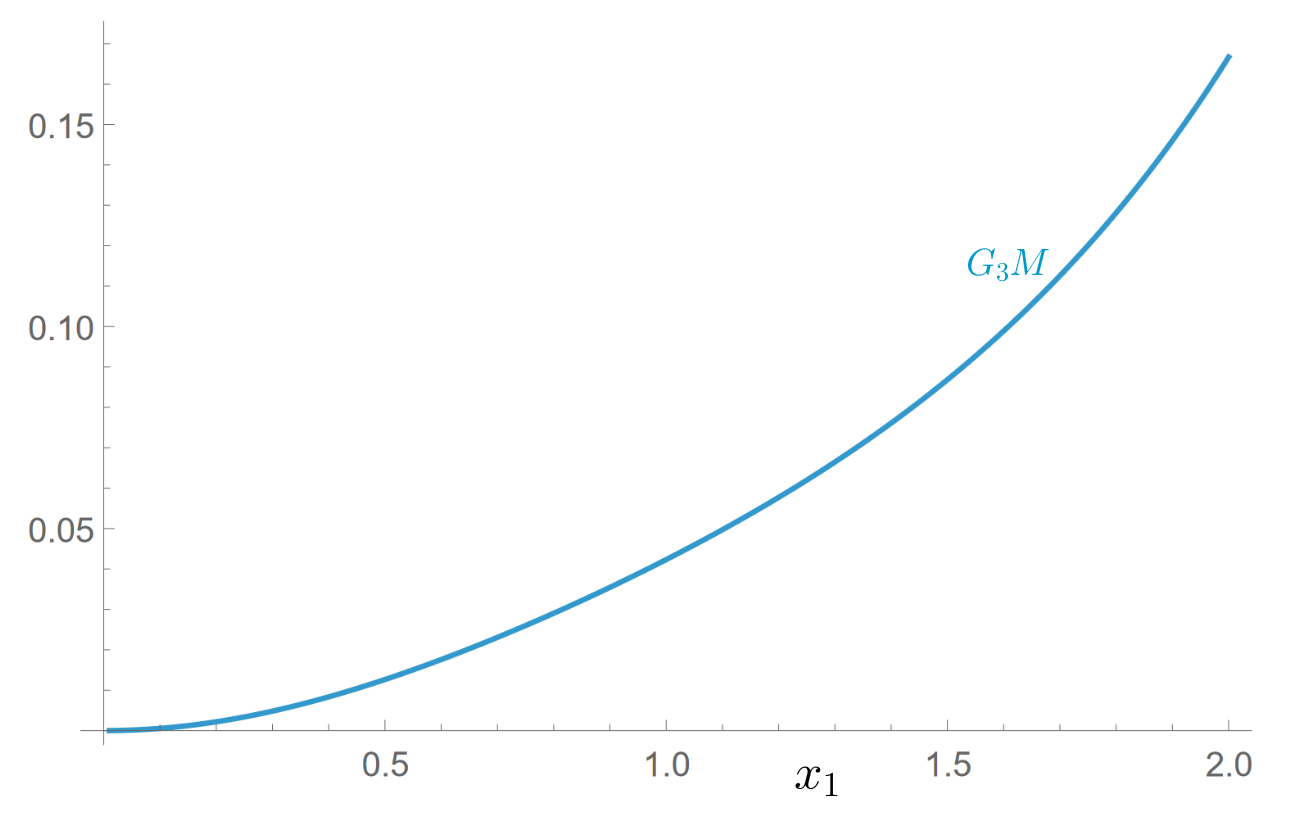}
\qquad
\includegraphics[width=0.5\textwidth]{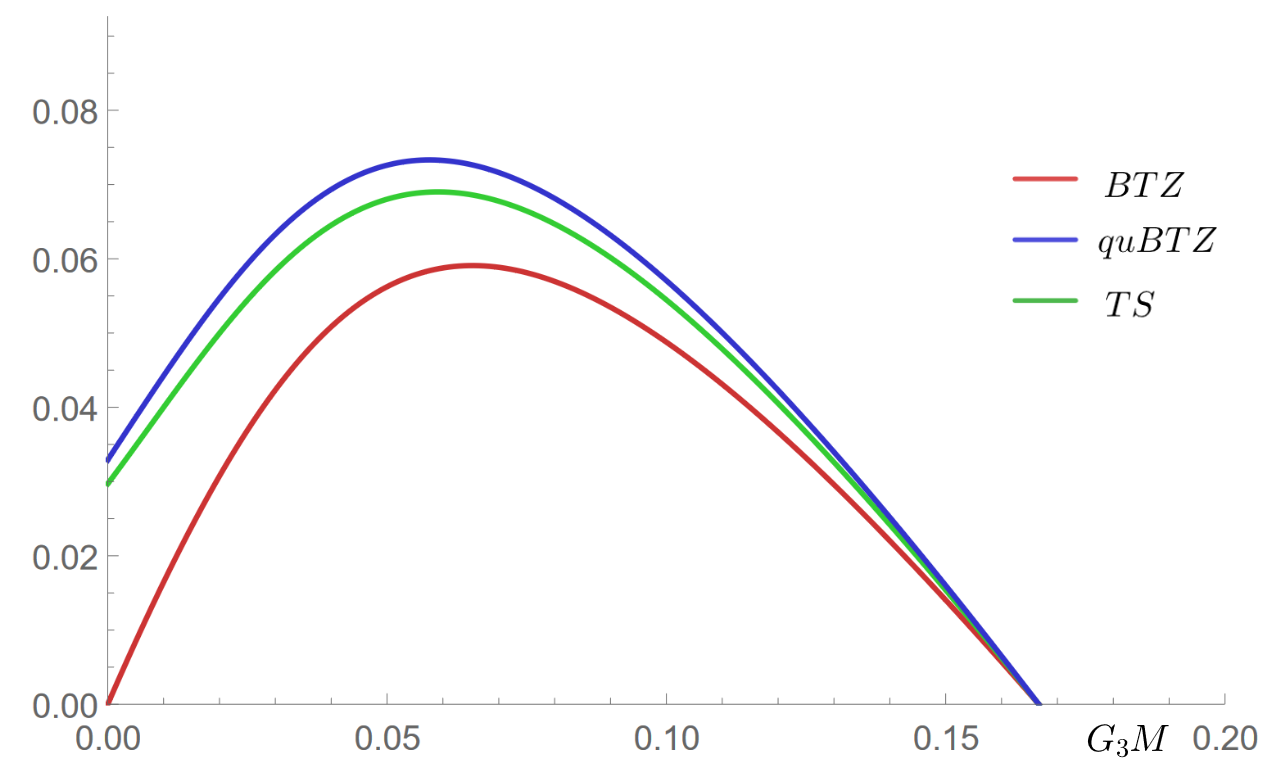}
\caption{
The late-time behaviors of rotating quBTZ and rotating BTZ, as well as $TS_{gen}$, for $\ell=0.1,a=1/4$ and $\ell_3=1$. The left figure shows that $M$ is increasing with $x_1$ and there is only one branch.  We can see up to $\mathcal{O}(1)$ quantum corrections, the three  quantities show the same behavior and all of them reach zero at the same point when the black hole becomes extremal.
}
\label{fig:vlmpot}
\end{figure}

In fact, there is a finer structure even for $M>0$. In Figure \ref{fig:volmass} and Fig. \ref{fig:vlmpot}, we take two different values of the parameter $a$ and show  the late-time growth rate of volume complexities of the rotating quBTZ and BTZ, as well as the quantity $TS_{gen}$.  Another subtlety is the dependence of mass $M$ \eqref{massang} of the rotating quBTZ on parameter $x_{1}$. When $a\ll \ell_3$, there is a range for $M$ to have two branches of $x_{1}$, see Figure \ref{fig:volmass}. Actually, the right figure of Fig. \ref{fig:volmass} is the zoom in of the right-upper corner of Fig. \ref{fig:volallval}.  While for $a\sim\ell_3$, for example, $a=1/4$, the function $M$ is an increasing function, see Figure \ref{fig:vlmpot}.


\section{Action Complexity}

In this section, we study the CA proposal for holographic complexity of the rotating quantum-corrected black hole. In particular, we are interested in the action complexity at constant $\bar t$ slice. As we explained in appendix B, the WdW patch of constant $t$ slice in old coordinates is ill defined, and so is its action complexity. Interestingly, it seems that the constant $\bar t$ slice is the only choice whose action complexity can be studied. In this case, the complexity  is given by  the on-shell action of the WdW patch in the four-dimensional bulk
\begin{equation}
    C_A=\frac{I(\mathcal{W}_{\bar t})}{\pi}.
\end{equation}
According to \cite{Brown:2015bva}, the action is given by 
\begin{equation}
        I({\mathcal{W}_{\bar t}})=I_{EH}+I_{brane}+I_{GHY}+I_{joints}+I_\kappa+I_{ct}
\end{equation}
where 
\bea
I_{EH}&=&\frac{1}{16\pi G_4}\int_{\mathcal{W}_{\bar t}} d^4x\sqrt{-g}(R+\frac{6}{\ell_4^2}),\\
I_{brane}&=&\frac{1}{2\pi G_4\ell}\int_{w_{\bar t}}d^3y\sqrt{-h},\\
I_{GHY}&=&\frac{1}{8\pi G_4}\int_{\mathrm{regulator}}d^3y\sqrt{|h|}K,\\
I_{joints}&=&\frac{1}{8\pi G_4}\int_{\mathrm{joints}} d^2y\sqrt{\sigma}\alpha,\\
I_\kappa &=&\frac{1}{8\pi G_4}\int_{\mathcal{N}}d\lambda d^2y\sqrt{\gamma}\kappa,\\
I_{ct}&=&\frac{1}{8\pi G_4}\int_{\mathcal{N}}d\lambda d^2y\sqrt{\gamma}\Theta\log(L_{ct}\Theta).
\eea
it includes several terms: $I_{EH}$, the Einstein- Hilbert (EH) term  with a negative cosmological constant  $\Lambda_{4}=-3/l_{4}^{2}$, where the integration domain $\mathcal{W}_{\bar t}$ is the bulk WdW patch; $I_{brane}$, the tensional brane action, where $w_{\bar t}=\mathcal{W}_{\bar t}\cap$ \textbf{brane} is the intersection between the WdW patch and the brane, with $h$ being the induced metric on the brane; $I_{GHY}$, the Gibbons-Hawking-York (GHY) term  defined on the regulator surface near AdS boundary; $I_{joints}$, the joint terms evaluated at the intersections of the null boundaries of the WdW
patch with other smooth hypersurfaces (which will be specified below), where $\alpha$ is some function that depends on the ‘boost angle’ between the normals to the hypersurfaces; $I_\kappa$, the null boundary terms defined on the null boundaries of the WdW patch  $\mathcal{N}$, which would vanish when the null surface is affinely parameterized; $I_{ct}$, the counterterm defined {also} on $\mathcal{N}$ which could be expressed in terms of its expansion {parameter} $\Theta$. The counterterm $I_{ct}$, which was first introduced in \cite{Lehner:2016vdi}, is necessary to remove the intrinsic ambiguity in the parametrization of the WdW null boundaries and make the action additive, but it introduces an arbitrary length scale $L_{ct}$.
 
\subsection{Regularized WdW patch}

The WdW patch $\mathcal{W}_{\bar{t}}$ is defined as the {AdS} bulk causal domain of dependence $D(\Sigma_{\bar{t}})$ of any spacelike hypersurface $\Sigma_{\bar{t}}$ anchored to the {asymptotic} boundary slice. In our case $\Sigma_{\bar{t}}$ is chosen to be the constant time hypersurface. Since the true asymptotic boundary locates at $r=-x/\ell$, where $r$ is $x$-dependent rather than being a constant, the shape of $\mathcal{W}_{\bar{t}}$ would be complicated due to the formation of caustics.

\begin{figure}[h]
\centering
\includegraphics[scale=0.9]{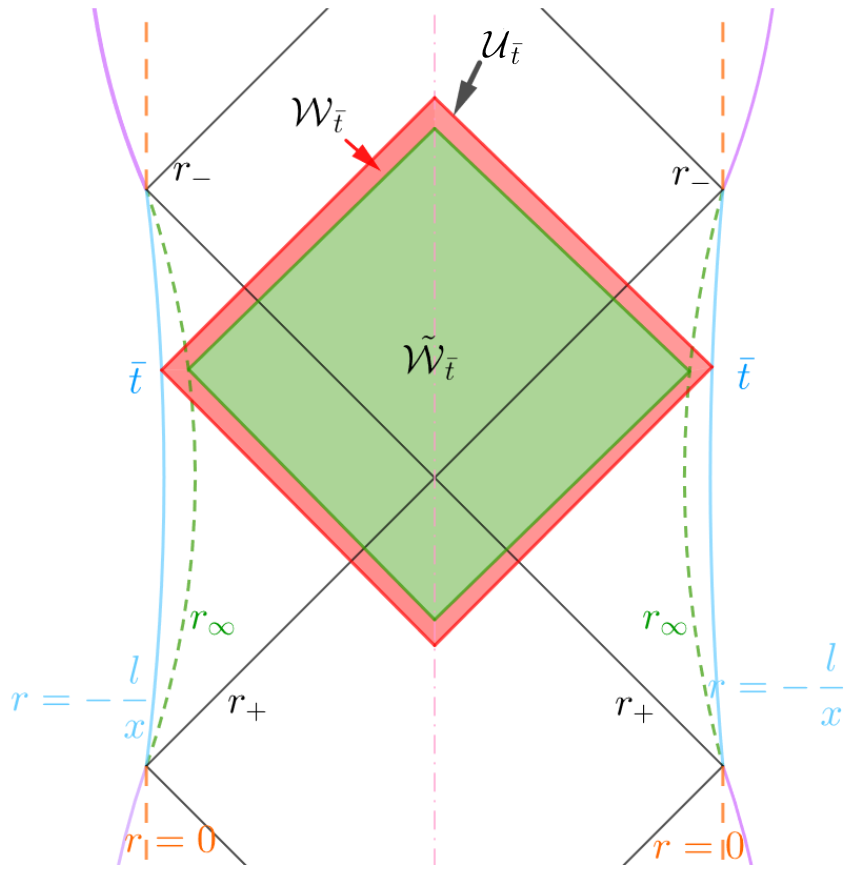}
\caption{The figure illustrates different WdW patches at a constant $(x,\bar{\phi})$ section of the spacetime. The true WdW patch $\mathcal{W}_{\bar t}$ lies between the red diamond $\mathcal{U}_{\bar t}$ starting from blue asymptotic boundary and green diamond $\tilde{\mathcal{W}}_{\bar t}$ starting from the cutoff surface $r=r_{\infty}$. }
\label{fig:actiontwo}
\end{figure}

Following the idea in the non-rotating  case \cite{Emparan:2021hyr}, we will use the regularized version of WdW patch $\tilde{\mathcal{W}}_{\bar t} \; { \subseteq \mathcal{W}_{\bar t} }  $ defined as the bulk causal domain of dependence of constant $\bar t$ hypersurface anchored to $r=r_\infty\to\infty$, and evaluate its  action. Remember that the true boundary locates at $x r=-\ell $ and the bulk spacetime extends beyond $r=+\infty$ into negative values of $r$. In order to obtain a finite action, we  need to regularize the WdW patch by introducing a cutoff hypersurface at $r=r_\infty$, then we let the regularized WdW patch $\tilde{\mathcal{W}}_{\bar t}$ start from this regulator surface at time $\bar t$, see Figure \ref{fig:actiontwo}. Consequently, the boundary of the WdW patch includes four null surfaces. Note that unlike the static case in \cite{Emparan:2021hyr}, we do not need to introduce a regulating surface at the singularity $r=0$ because that the regularized WdW patch ($\tilde{\mathcal{W}}_{\bar t}$) $\mathcal{W}_{\bar t}$ never goes into the inner horizon and so avoid the singularity.

The action complexity associated to  this regularized WdW patch is denoted as $\tilde C_A(\bar t)$ 
\begin{equation}
    \tilde C_A(\bar t)=\frac{I(\tilde{\mathcal{W}}_{\bar t})}{\pi}={C_A(\bar t)-C_{UV}}.
\end{equation}
According to \cite{Emparan:2021hyr}, the difference between $C_A(\bar t)$ and $\tilde{C}_A(\bar t)$ could be interpreted as the UV part of the complexity, {which is} associated to the short-range correlations in the state of the dCFT system. We will see later that the late-time growth of the action is dominated by the regularized action $\tilde C_A(\bar t)$.

The boundary of $\tilde{\mathcal{W}}_{\bar t}$ is determined by emitting future and past null congruences from a bulk finite cutoff surface at constant $\bar{t}$ and $r=r_{\infty}<+\infty$. From general analysis on the spacetime metric, it is not hard to see that its shape is like a ``causal diamond". We may define a new deformed causal diamond $\mathcal{U}_{\bar t}$  by following radial null geodesics from the real asymptotic boundary at $r=-\ell/x$. The real WdW patch $\mathcal{W}_{\bar t}$ only involves the shortest  null geodesics, not simple radial null geodesics, so generally its boundary would move along the transverse coordinates $x$ and $\bar{\phi}$. Therefore, it would satisfy $\tilde{\mathcal{W}}_{\bar t} \subseteq \mathcal{W}_{\bar t} \subseteq \mathcal{U}_{\bar t}$, as shown in Fig. \ref{fig:actiontwo}. At the late-time regime, these three bulk codimension-zero regions would coincide.

The boundary null hypersurfaces of WdW patch is constrained by an equation $\Phi(x) =$ constant with null normal satisfying the condition $g^{\mu\nu}\partial_\mu\Phi\partial_\nu\Phi=0$.  Due to the axial symmetry, we define Eddington-Finkelstein-like coordinates as
\begin{equation}
    v=\bar t+r^*(r,x),\quad u=\bar t-r^*(r,x)
\end{equation}
and let the null hypersurfaces be given by $u=$ const and $v=$ const. In both cases, the
null condition reads
\begin{equation}\label{nullWDW}
    \Delta_r(\partial_rr^*)^2+\Delta_x(\partial_xr^*)^2=\frac{1}{(1-\tilde a^2)^2\Delta^2}\Big[\frac{(r^2+a^2x_1^2)^2}{\Delta_r}-\frac{a^2(x^2-x_1^2)^2}{\Delta_x}\Big].
\end{equation}
The right-hand side is positive for $r\geq r_+$. To solve this partial differential equation for $r^*$, we 
follow the procedure used in \cite{AlBalushi:2019obu} by introducing one auxiliary function $\zeta(r,x)$ to define
\begin{equation}
Q^2=\frac{1}{(1-\tilde a^2)^2\Delta^2}[(r^2+a^2x_1^2)^2-a^2\zeta \Delta_r],\quad P^2=\frac{1}{(1-\tilde a^2)^2\Delta^2}a^2[\zeta\Delta_x-(x^2-x_1^2)^2].
\end{equation}
Then the equation \eqref{nullWDW} is  solved by
\begin{equation}
    \partial_rr^*=\frac{Q}{\Delta_r},\quad \partial_xr^*=\frac{P}{\Delta_x}.
\end{equation}
Alternatively, the solution to \eqref{nullWDW} is given by solving the exact integral
\begin{equation}\label{dr*}
    dr^*=\frac{Q}{\Delta_r}dr+\frac{P}{\Delta_x}dx.
\end{equation}
The consistency equation $d^2r^*=0$  implies that the auxiliary  function $\zeta$ must have a differential of the form
\begin{equation}\label{dzeta}
    d\zeta=\frac{1}{s}\Big(-\frac{dr}{Q}+\frac{dx}{P}\Big)
\end{equation}
for some function $s=s(r,x)$. 
The degenerate intrinsic metric on the null hypersurface can be obtained by substituting $du=0$ or $dv=0$, together with \eqref{dr*} and \eqref{dzeta} into the metric \eqref{metric-can}. 
Then we find
\begin{equation}
 ds^2\Big|_{\partial\mathcal{N}}=\frac{\ell^2}{(\ell+xr)^2}\Big[\frac{\Delta^2(1-\tilde a^2)^2s^2P^2Q^2\rho^2}{\bar{\Sigma}^2}d\zeta^2+\Phi^2(d\bar\phi-wd\bar t)^2\Big].
\end{equation}
The determinant of the induced metric on the null boundary of the WdW patch is given by
\begin{equation}\label{det}
    \sqrt{\gamma}=\frac{\ell^2(1-\tilde a^2)\Delta^2 sPQ}{(\ell+xr)^2}.
\end{equation}
In computing the action complexity, the presence of caustics, which is signalled by the vanishing of above determinant, is fatal. We prove the absence of caustics in the WdW patch of rotating quBTZ in Appendix A.

\subsection{Evaluating the action}

Now we are ready to evaluate the action complexity.
The null normal one-forms associated to the null hypersurfaces  can be obtained. Focusing on the future part of the WdW patch and choosing the null normals to be outward directed from the boundary of the WdW patch, we get
\begin{equation}
    k_v= \alpha_1dv=\alpha_1(1,\frac{Q}{\Delta_r},\frac{P}{\Delta_x},0), \quad k_u=-\alpha_2du=\alpha_2(-1,\frac{Q}{\Delta_r},\frac{P}{\Delta_x},0)
\end{equation}
with $\alpha_1,\alpha_2$ being two positive constants.
The inner product between two null-normal 1-forms is given by
\begin{equation}\label{norm-kukv}
    |k_v\cdot k_u|=| \alpha_1\alpha_2\frac{(\ell+xr)^2}{\ell^2}\Big(\frac{1}{N^2}+\frac{Q^2}{\Delta_r\rho^2}+\frac{P^2}{\Delta_x\rho^2}\Big)|=|2 \alpha_1\alpha_2\frac{(\ell+xr)^2}{\ell^2N^2}|.
\end{equation}
Moreover, it can be checked directly that both $k_v$ and $k_u$ are affinely parametrized, i.e. $k^\mu_a\nabla_\mu k^\nu_a=0$ for $a=u,v$ and hence the action $I_\kappa=0$.

At the intersection of two null surfaces, we  have the relation $dt=dr^*=0$ and in particular, the vanishing of $dr^*$ leads to a relation between $dr$ and $dx$, 
\begin{equation}
   dr^*=\frac{Q}{\Delta_r}dr+\frac{P}{\Delta_x}dx=0.
\end{equation}
Therefore, \eqref{dzeta} becomes
\begin{equation}
    d\zeta=\frac{1}{s}\Big(\frac{1}{P}+\frac{P\Delta_r}{\Delta_xQ^2}\Big)dx=\frac{\bar{\Sigma}^2}{(1-\tilde a^2)^2\Delta^2sPQ^2\Delta_x}dx.
    \end{equation}
Using this relation, the volume form of the joint surface is given by
\begin{equation}\label{int-jnt}
\sqrt{\gamma}d\zeta d\phi=dxd\phi\frac{\ell^2\bar{\Sigma}^2}{(1-\tilde a^2)(\ell+xr)^2\Delta_xQ}.
\end{equation}

\begin{figure}[h]
\centering
\includegraphics[scale=0.9]{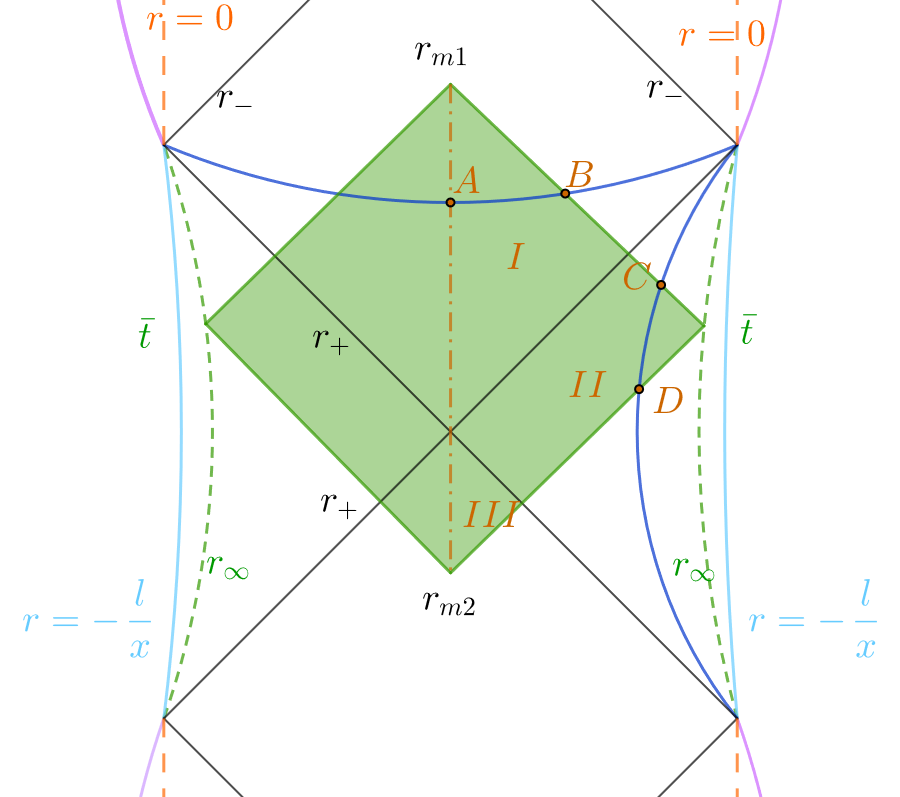}
\caption{Regularized WdW patch $ \Tilde{\mathcal{W}}_{\bar{t}}$ for constant $(x,\bar{\phi})$ section. The radius of upper and lower tips $r_{m1}(\bar{t})$ and $r_{m2}(\bar{t})$ is implicitly determined by \eqref{rm1} and \eqref{rm2}.  The points $A$ and $B$ are located on the constant-$r$ surface with the coordinates $(0,r)$ and $(\bar t+r^{*}(r_\infty)-r^*(r),r)$; The points $C$ and $D$ are still located on the constant-$r$ surface with the coordinates $(\bar t+r^{*}(r_\infty)-r^*(r),r)$ and $(\bar t-r^{*}(r_\infty)+r^*(r),r)$. }
\label{fig:calstru}
\end{figure}

At each constant-$x$ slice, we denote  the future-meeting point of two null surfaces as $m_1$ and  the past-meeting point as $m_2$.  For the symmetric configuration $\bar t_L=\bar t_R=\bar t$, we have
\begin{equation}\label{rm2}
 r^*(r_{m_2},x)=r^*(r_\infty,x)-\bar t.
\end{equation}
Taking derivative with respect to $\bar t$ and using \eqref{dr*} gives
\begin{equation}
   \frac{Q}{\Delta_r}\bigg \vert_{{r=r_{m2}}} \frac{\partial r_{m_2}}{\partial \bar t}=-1.
\end{equation}
Similarly, for $r_{m_1}$, we have
\begin{equation}\label{rm1}
    r^*(r_{m_1},x)=r^*(r_{{\infty}},x)+\bar t,\quad  \frac{Q}{\Delta_r}\frac{\partial r_{m_1}}{\partial\bar t}=1.
\end{equation}

\paragraph{Bulk contribution}
Similar to the non-rotating case, the existence of the brane and the junction condition will affect the curvature, and as a result the bulk solution \eqref{metric-can} satisfies the inhomogeneous Einstein equation with a source localized on the brane. More specifically, we have
\begin{equation}\label{eqn-rdd}
    R_{\mu\nu}-\frac{1}{2}Rg_{\mu\nu}+\Lambda_4g_{\mu\nu}=-\frac{4}{\ell}h_{\mu\nu}\delta(X),
\end{equation}
where $X$ is a normal parameter to the brane that grows towards $x > 0$ and satisfies $dX/dx = 1/r$ at $x = 0$. Taking the trace of \eqref{eqn-rdd}, we find
\begin{equation}
    R-2\Lambda_4=-\frac{6}{\ell_4^2}+\frac{12\delta(x)}{\ell r}.
\end{equation}
This last term is proportional to $\delta(x)$ and becomes an integral over $w_t$ when being substituted into the Einstein-Hilbert action. Then we have
\begin{equation}
    I_{EH}=-\frac{3\mbox{Vol}(W_{\bar t})}{8\pi G_4\ell_4^2}+\frac{3\mbox{Vol}(w_{\bar t})}{4\pi G_4\ell}.
\end{equation}
Incorporating the bare brane contribution $I_{brane}$ into the total bulk contribution, we have
\begin{equation}
    I_{bulk}=I_{EH}+I_{brane}=-\frac{3\mbox{Vol}(W_{\bar t})}{8\pi G_4\ell_4^2}+\frac{\mbox{Vol}(w_{\bar t})}{4\pi G_4\ell}.
\end{equation}
To evaluate the bulk action, it is also convenient to divide the whole  WdW patch into three regions I, II, and III, as shown in Figure \ref{fig:calstru}.
Performing the corresponding integration over the $t$-coordinate for each of the regions yields
\begin{equation}
    \begin{aligned}
       &\mbox{Vol}(I)=\int^{2\pi}_0d\bar\phi\int_0^{x_1}dx\int_{r_{m_1}}^{r_+}dr\frac{4\Delta^2(1-\tilde a^2)\ell^4(r^2+a^2x^2)}{(\ell+xr)^4}(\bar t+r_*(r_\infty)-r_*(r)),\\
       &\mbox{Vol}(II)=-\int^{2\pi}_0d\bar\phi\int_0^{x_1}dx\int^{r_\infty}_{r_+}dr\frac{{8} \Delta^2(1-\tilde a^2)\ell^4(r^2+a^2x^2)}{(\ell+xr)^4}(r_*(r)-r_*(r_\infty)),\\
       &\mbox{Vol}(III)=\int^{2\pi}_0d\bar\phi\int_0^{x_1}dx\int_{r_{m_2}}^{r_+}dr\frac{4\Delta^2(1-\tilde a^2)\ell^4(r^2+a^2x^2)}{(\ell+xr)^4}(-\bar t+r_*(r_\infty)-r_*(r)),
    \end{aligned}
\end{equation}
where we have multiplied a factor 2 to account for the two-sided spacetime.
Therefore, we have
\begin{equation}
    \mbox{Vol}(W_{\bar t})={V_{0}}+ 8\pi (1-\tilde a^2)\Delta^2\ell^4\int^{x_1}_0dx\int^{r_{m_2}}_{r_{m_1}}dr\frac{(r^2+a^2x^2)}{(\ell+xr)^4} \( \bar{t}   {-r_{*}(r_{\infty})+r_{*}(r) ) } \),
\end{equation}
where {$V_{0}$ is still an time-dependent term which is different from the one of the static quBTZ case
\be 
V_{0}=16 \pi (1-\tilde a^2)\Delta^2\ell^4\int^{x_1}_0dx\int^{r_{\infty}}_{r_{m_1}}dr\frac{(r^2+a^2x^2)}{(\ell+xr)^4} \( r_{*}(r_{\infty})-r_{*}(r)   \).
\ee} 
Repeating the previous analysis for the volume on the brane, we have
\begin{equation}
    \mbox{Vol}(w_t)=  {v_{0}} + 4\pi (1-\tilde a^2)\Delta^2\int^{r_{m_2}}_{r_{m_1}}dr r \( \bar t  {- r_{*}(r_{\infty})+r_{*}(r) }\),
\end{equation}
 where $v_{0}$ is a time-dependent term 
\be 
v_{0}=8\pi (1-\tilde a^2)\Delta^2\int^{r_{\infty}}_{r_{m_1}}dr r
\( r_{*}(r_{\infty})-r_{*}(r) \).
\ee

\begin{figure}[htbp]
\centering
\includegraphics[width=0.46\textwidth]{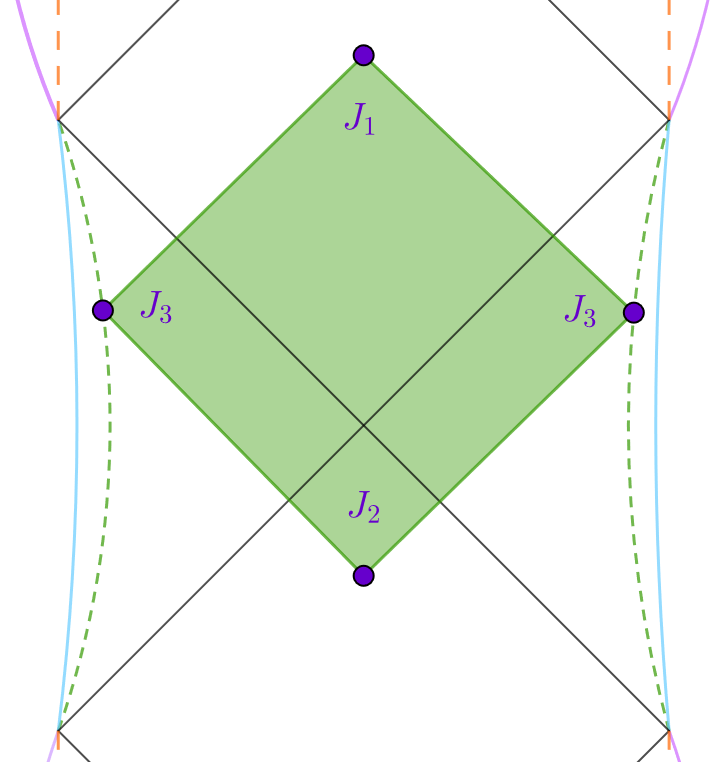}
\qquad
\includegraphics[width=0.46\textwidth]{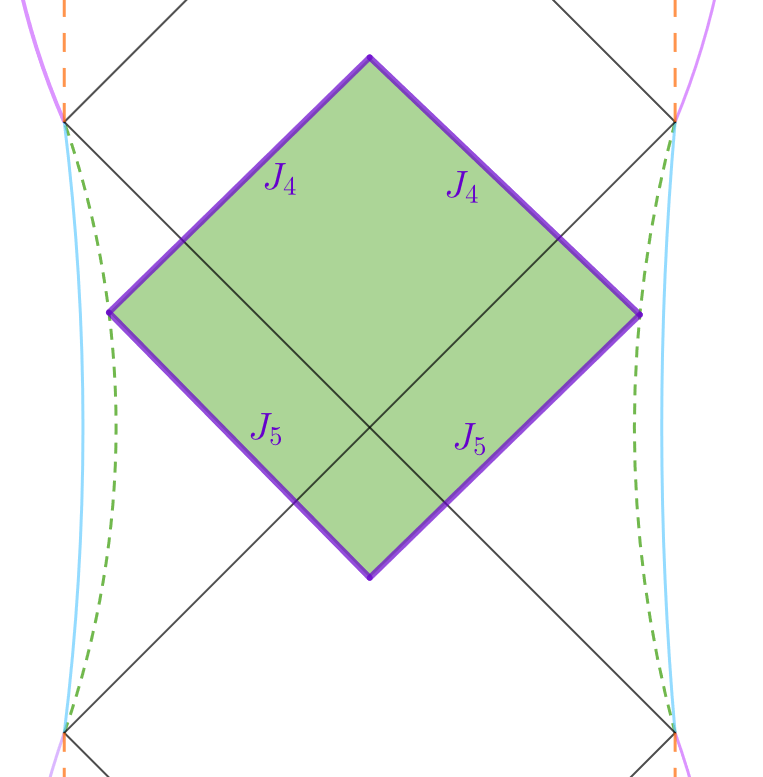}
\caption{\textbf{Left}: Purple dots represent the standard joint surface in a constant $(x,\phi)$ section of the geometry. \textbf{Right}: Purple lines represent the new joint surfaces in a constant $\phi$ section on the brane.   }
\label{fig:joints}
\end{figure}

\paragraph{Contributions from the joints}
As shown in the left figure of Fig. \ref{fig:joints}, there are three kinds of  joint terms which are defined at the future-intersection surface $J_1$,  at the past-intersection surface $J_2$, and at the cutoff surface $J_3$ respectively.

To compute the joint terms at $J_1$ and $J_2$, 
we note that $r_{m_1}$ and $r_{m_2}$ are inside the outer horizon but outside the inner horizon, i.e. $r_-<r_{m_i}<r_+$. Therefore, $H(r_{m_i})<0$.    
The specific form of the joint action is given by
\begin{equation}
    I_{jnt}=\frac{1}{8\pi G_4}\int_{\mathrm{joints}}d^2y\sqrt{\sigma}\log\Big|\frac{k_u\cdot k_v}{2}\Big|.
\end{equation}
Substituting \eqref{norm-kukv} and \eqref{int-jnt} into this joint action, we get
\begin{equation}\label{Ijnt}
    I^i_{jnt}=-\frac{1}{4\pi G_4(1-\tilde a^2)}\int^{2\pi}_0d\bar\phi\int^{x_1}_0dx\frac{\ell^2\bar\Sigma^2}{(\ell+xr_{m_i})^2\Delta_xQ}\log\Big(\frac{|\ell^2N^2|}{|\alpha_1\alpha_2(\ell+xr_{m_i})^2|}\Big)
\end{equation}
with $i=1,2$ labelling the joint surfaces at $r_{m_i}$. Finally, the joint terms at $J_3$ are similarly given by \eqref{Ijnt} but with $r_{m_i}$ being replaced by $r_\infty$.

\paragraph{Contributions from new terms}
In our setup, since we glue two identical copies across the brane, new joint surfaces arise as the intersection between the two sides of the null boundary $J_4,J_5$ as shown in Fig. \ref{fig:joints}. As we discussed in appendix A, the null normal vectors is tangent to the brane on both sides of the brane. From the junction conditions, the tangent properties of the spacetime remain continuous across the brane, and so do these normal vectors. Consequently, the normals are $n^\mu_1=n^\nu_2\propto (dt)^\mu\pm\frac{Q}{\Delta_r}(dr)^\mu$ where $1,2$ label the normals on two sides. As a result, the boost angle in the integrand of the joint action vanishes so the new joint surfaces give no contributions.

\paragraph{Contributions from the counter term}
Finally let us consider the counter-term action
\begin{equation}
    I_{ct}=\frac{1}{8\pi G_4}\int_{\partial\mathcal{N}}d\lambda d^2y\sqrt{\gamma}\Theta\log(L_{ct}\Theta)
\end{equation}
{where $L_{ct}$ is an arbitrary length scale and} $\Theta=\partial_\lambda{\log}\sqrt{\gamma}$. We will focus on the right future-boundary of the WdW patch $v=const.$ The integration runs over $\zeta,\phi$ and the null coordinate $\lambda$ which satisfies
\begin{equation}
    k_v^\mu=\frac{\partial x^\mu}{\partial\lambda}=\alpha_2\frac{(\ell+xr)^2}{\ell^2}\Big(\frac{-1}{N^2},\frac{Q}{\rho^2},\frac{P}{\rho^2},0\Big).
\end{equation}
Introducing an auxiliary null vector $N^\mu$ such that $N\cdot k_v=-1$, 
\begin{equation}
  \hat N=\frac{-k_u}{k_u\cdot k_v}=\frac{\ell^2N^2}{2\alpha_1\alpha_2(\ell+xr)^2}k_u,
\end{equation}
we have
\begin{equation}
     d\lambda=-N_\mu dx^\mu=-\frac{\ell^2N^2}{2\alpha_2(\ell+xr)^2}(dt-dr^*)=\frac{\ell^2N^2}{\alpha_2(\ell+xr)^2}dr^*
\end{equation}
where we have used the fact that $dv=dt+dr^*=0$. Besides, combining \eqref{dr*} with \eqref{dzeta} and cancelling $dr$, we find
\begin{equation}
    d\zeta=\frac{1}{sQ^2}\Big(-\Delta_rdr^*+\frac{\bar\Sigma^2}{\Delta^2(1-\tilde a^2)^2P\Delta_x}dx\Big).
\end{equation}
Then we obtain the counter-term action given by
\begin{equation}\label{counter-act}
 I_{ct}=   \frac{1}{8\pi G_4}\int d\lambda d^2y\sqrt{\gamma}\Theta\log(l_{ct}\Theta)=\frac{1}{4G_4}\int dr^*dx\frac{\ell^4(1-\tilde a^2)\rho^2\Delta_r}{\alpha_2(\ell+xr)^4Q}\Theta\log(l_{ct}\Theta).
\end{equation}

\subsection{Late-time dependence}

In the late-time limit $\bar t\to\infty$, by virtue of \eqref{rm2} and \eqref{rm1}, we have
\begin{equation}\label{late-r}
   \lim_{\bar t\to\infty} r_{m_2}= r_+,\quad \lim_{\bar t\to\infty} r_{m_1}= r_-.
\end{equation}
As a result, the late-time limit of the functions involved in the computations is
\begin{equation}\label{latetime}
    \lim_{\bar t\to\infty}\Delta_r=\lim_{\bar t\to\infty}N^2=0,~~\lim_{\bar t\to\infty}\bar\Sigma^2=(r_\pm^2+a^2x_1^2)^2\Delta_x,~~\lim_{\bar t\to\infty}Q=\frac{r_\pm^2+a^2x_1^2}{\Delta(1-\tilde a^2)}
\end{equation}
which further implies that the derivative of the joint radius with respect to time approaches to 0, 
\begin{equation}
\lim_{\bar t\to0}\frac{\partial r_{m_i}}{\partial\bar t}=0.
\end{equation}
The late-time growth rate of the bulk action is then given by
\begin{equation}
    \begin{aligned}\label{Ibulk:dt}
       \frac{dI_{bulk}}{d\bar t}&=\frac{-3(1-\tilde a^2)\Delta^2\ell^4}{G_4\ell_4^2}\int^{x_1}_0dx\int^{r_+}_{r_-}dr\frac{r^2+a^2x^2}{(\ell+xr)^4}+\frac{(1-\tilde a^2)\Delta^2}{2G_4\ell}(r_+^2-r_-^2)\\
      & =\frac{(1-\tilde a^2)\Delta^2\ell(r_-^2-r_+^2)}{2G_4}\Big(\frac{1}{\ell_3^2}+\frac{a^2}{r_-^2r_+^2}\Big)\Big(1-\frac{2\ell\ell_3^2(a^2r_-r_+x_1^2+a^2r_+^2x_1^2+r_-^2r_+^2+a^2r_-^2x_1^2)}{x_1^3r_-r_+(r_-+r_+)(r_-^2r_+^2+a^2\ell_3^2)}\Big),
    \end{aligned}
\end{equation}
where we have used $\ell_4=\big(1/\ell^2+1/\ell_3^2\big)^{-1/2}$ and performed the  expansion in $\ell$ up to linear order. It is easy to see that the  joint action at the cutoff surface $J_3$ is time independent.
The time-dependence of the joint action comes from the time derivative of the integrand in \eqref{Ijnt}. In the late-time limit the only terms that survive are those where the derivative acts on the $\Delta_r$ factor contained within the lapse function $N$. Using the explicit expressions, we find 
\begin{equation} 
    \begin{aligned}
     \lim_{\bar t\to\infty}  \frac{dI_{jnt}^i}{d\bar t}&=-\frac{\Delta}{2 G_4}\int^{x_1}_0dx\frac{\ell^2(r^2_{m_i}+a^2x_1^2)}{(\ell+xr_{m_i})^2}\frac{\Delta_r'}{\Delta_r}\frac{dr_{m_i}}{d\bar t}\\
       &=\frac{(-1)^i\ell x_1r_{m_i}^2\Delta^2(1-\tilde a^2) H'}{2G_4(\ell+x_1r_{m_i})}.
    \end{aligned}\label{timedepjoint}
\end{equation}
Finally, for the counter-term action\eqref{counter-act}, the time dependence is implicitly contained in the extremum of integration, corresponding to the tip of the WdW patch, i.e., in $r^*$. Taking the time derivative of \eqref{counter-act},
we obtain
\begin{equation}
    \frac{dI_{ct}}{d\bar t}=\frac{1}{4G_4}\int dx\frac{\ell^4(1-\tilde a^2)\rho^2\Delta_r}{\alpha_2(\ell+xr)^4Q}\Theta\log(l_{ct}\Theta),
\end{equation}
which  gives vanishing contribution when evaluated at $r_+$ or $r_-$. In conclusion, we find that only the bulk action and the joint action have non-trivial  late-time dependence. 
Summing up these contributions, expanding the result in small $\ell$, and considering the case that $a\gg \ell$ and $x_1\gg (\ell/ \ell_3)^{1/3}$, we find
\begin{equation}\label{expand-AC}
    \lim_{\bar t\to\infty}\frac{dI}{d\bar t}=\frac{\Delta^2(1-\tilde a^2)\sqrt{\ell_3^2-4a^2}}{2G_3\ell_3}\Big(1+\frac{\ell}{\ell_3}F_A(x_1,\tilde a)\Big)
\end{equation}
where $F_A$ is a complicated function. The ratio $F_A/F_V$ or $F_A/F_{gen}$ is a finite number of order 1.  Intuitively, the leading term should be the late-time growth of the action complexity of classical BTZ black hole, and we show that this is indeed the case.
Firstly, note that using \eqref{effact} the bulk action of BTZ is given by
\be
I_{bulk}^{BTZ}=-\frac{\mathrm{Vol}(w_{\bar t})}{4\pi G_3\ell_3^2}.
\ee
Its late-time derivative is given by
\be
\lim_{\bar t\to\infty}\frac{dI_{bulk}^{BTZ}}{d\bar t}=-\frac{(1-\tilde a^2)\Delta^2(r_+^2-r_-^2)}{2G_3\ell_3^2}.
\ee
This is just the leading term of \eqref{Ibulk:dt} using $G_4\sim 2G_3\ell$ and $r_+r_-=a\ell_3+\mathcal{O}(\ell)$. For the joint action, it can be also checked that the leading part in \eqref{timedepjoint} gives the joint action for the BTZ black hole by direct computation. An easier way to see this is to note that with one $\ell$ cancelled against $G_4$, there is a remaining factor $\ell/(\ell+xr_{m_i})^2$ in the integrand of \eqref{timedepjoint}. Using similar argument as before, it can be shown that 
\be
\lim_{\ell\to0}\frac{\ell}{(\ell+xr_{m_i})^2}=\delta(xr_{m_{i}}).
\ee
Therefore, the leading contribution to \eqref{timedepjoint} comes from the brane $x=0$ which reduces to the joint action for BTZ automatically. To conclude, the leading term in \eqref{expand-AC} is exactly the late-time slope of complexity of BTZ black hole.

Let us comment on the validity of the computation of late-time dependence using the regularized WdW patch instead of the full WdW patch. The key property that makes this feasible is that at very late time,  the conditions \eqref{late-r} and \eqref{latetime} are always true no matter which choice we make. Therefore, the range of integration of the bulk action in \eqref{Ibulk:dt} remains unchanged when we use the full WdW patch. Besides, the time derivative of the joint action and the counter-term action are also the same since \eqref{latetime} holds. Therefore, we can use the regularized WdW patch to calculate the late-time dependence of the action complexity without changing the result.

From the above discussions, we see that the late-time slope of action complexity matches with that of volume complexity \eqref{CV:dt} up to a factor of 2 at the leading order, while the quantum corrections to them differ slightly. This result is very different from the case of  non-rotating quBTZ. In particular, the action complexity of non-rotating quBTZ is independent of $\ell$ so that no expansion can be performed. 
However,  the action complexity for rotating quBTZ shows explicitly that the quantum correction will appear at the linearized order of $\ell$.  This suggests that the late-time slope of the action complexity is discontinuous when we take the non-rotating limit, i.e.
\be
\lim_{a\to0}\lim_{\bar t\to\infty}\frac{dC_A}{d\bar t}\neq \lim_{\bar t\to\infty}\frac{dC_A^{a=0}}{d\bar t}.
\ee
To study this mismatch in detail, we consider taking the non-rotating limit $a\to0$ carefully in the next subsection.


\begin{figure}[htbp]
\centering
\includegraphics[width=0.47\textwidth]{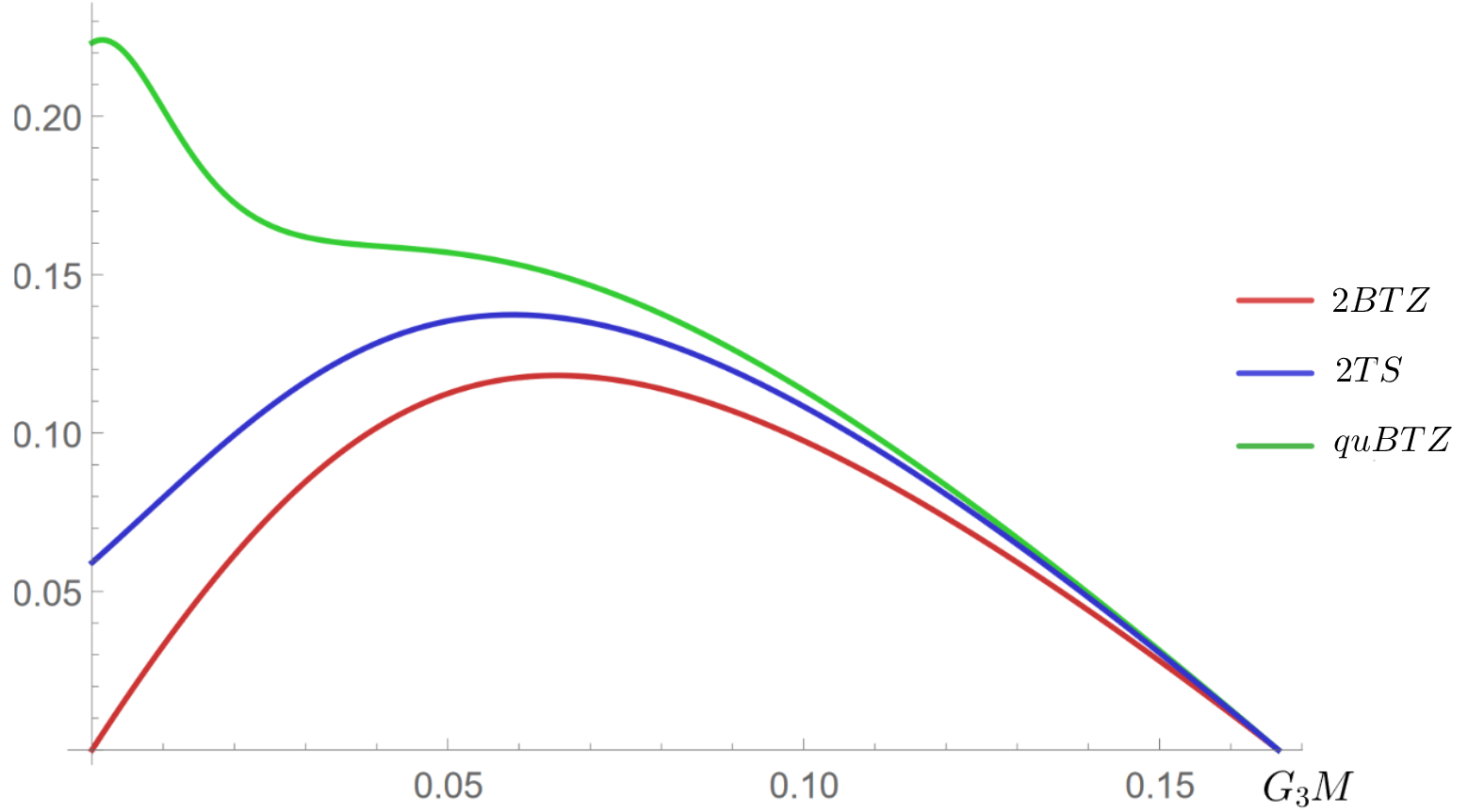}
\qquad
\includegraphics[width=0.44\textwidth]{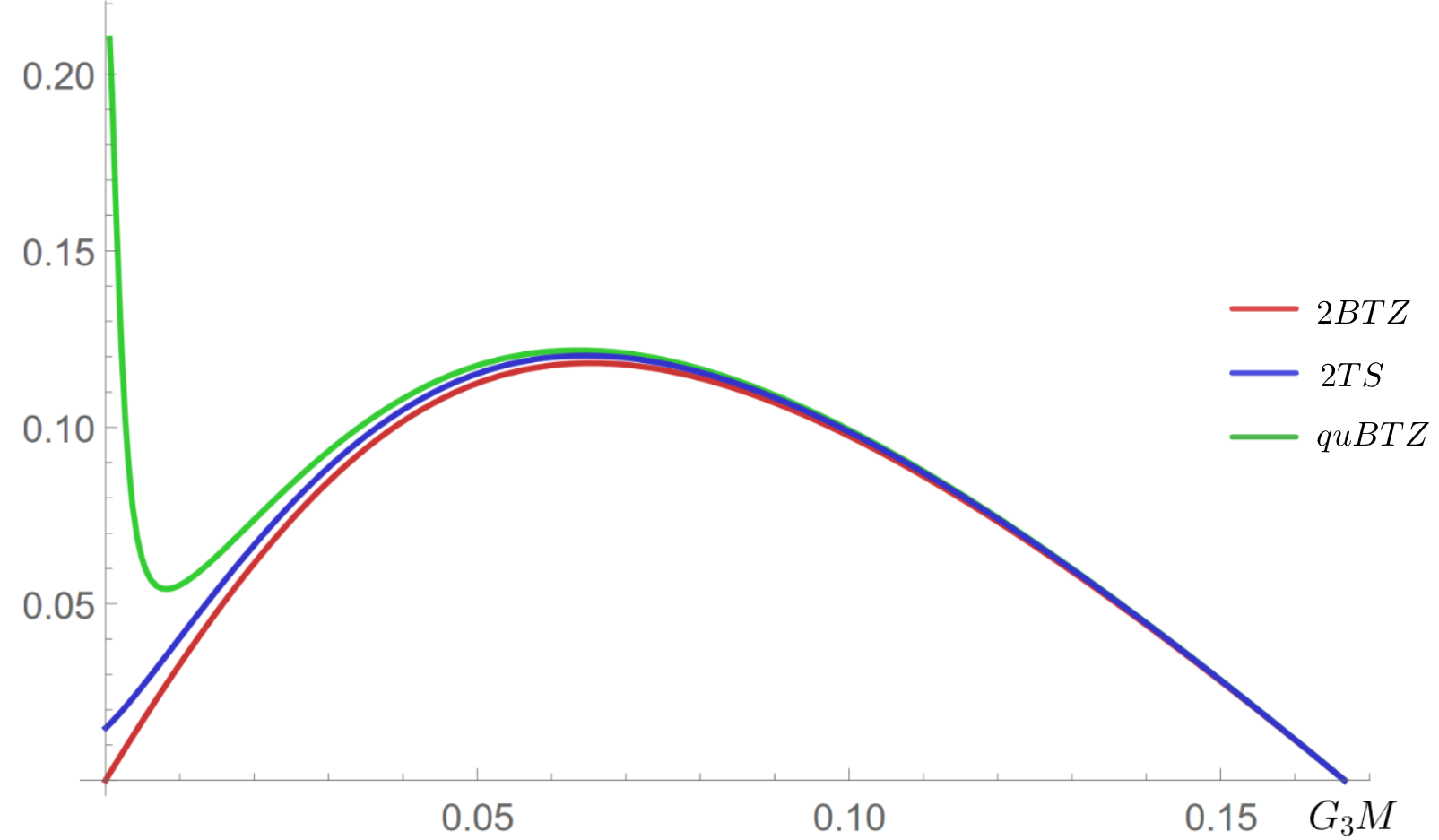}
\caption{The comparison of late-time growth rate of the action complexity for the rotating quBTZ, the one of the volume complexity for classical BTZ and $TS_{gen}$. The latter two quantities are both doubled. The parameters take values  $a=1/4$, $\ell=0.1,$ $\ell_3=1$ in left figure, and $a=1/4$, $\ell=0.01$, $\ell_3=1$ in right figure for  quBTZ. Apart from a narrow region around $x_{1}=0$, the three complexity-related quantities show better match with smaller $\ell$.   }
\label{fig:acttwotwo2}
\end{figure}

\begin{figure}[htbp]
\centering
\includegraphics[width=0.6\textwidth]{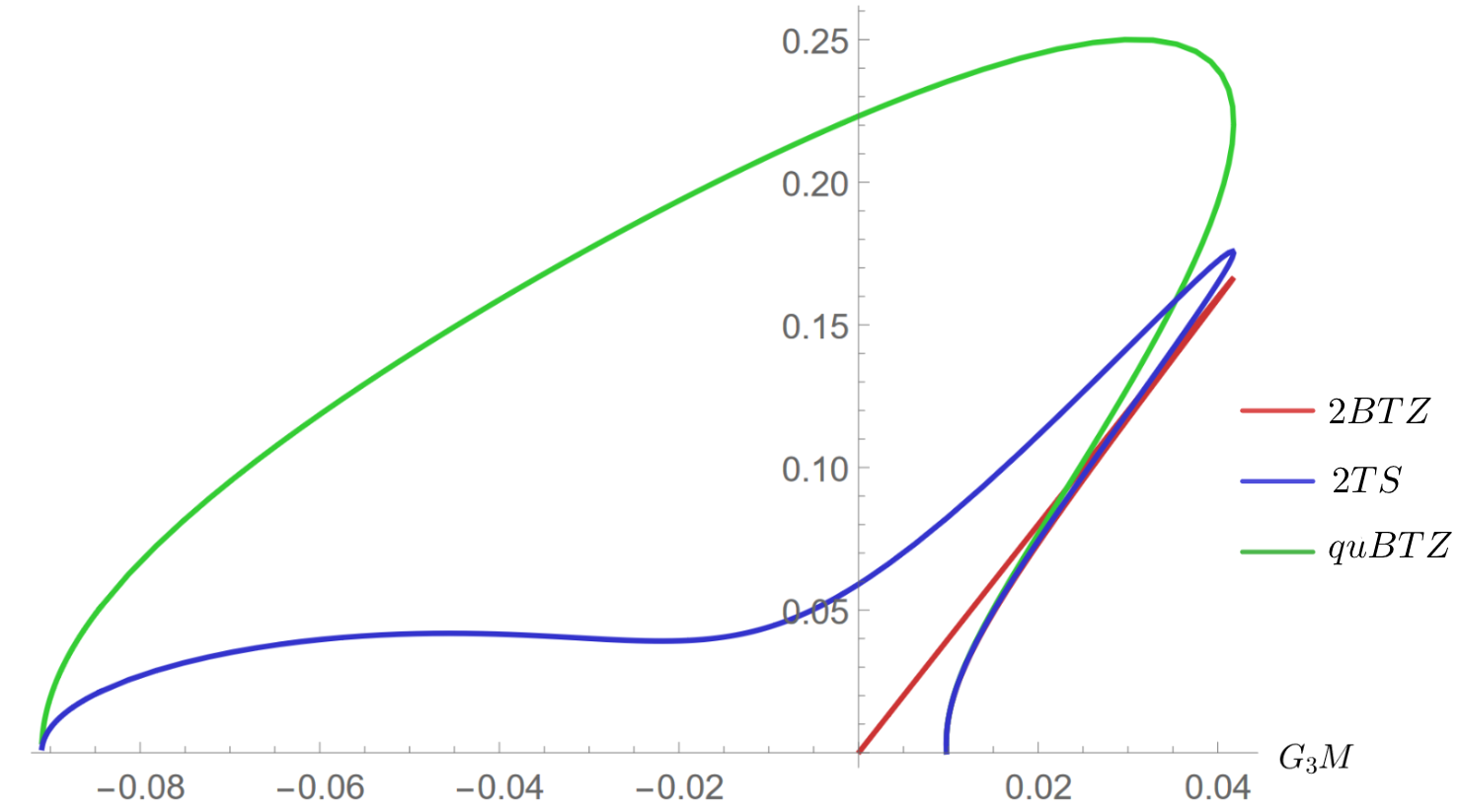}
\caption{The late-time growth rate of the action complexity for the rotating quBTZ compared with twice the value of thermodynamic quantity $TS_{gen}$(blue) and twice the late-time slope of volume complexity of classical BTZ(red) with the parameters $a=0.01$, $\ell=0.1$ and $\ell_3=1$. We can see that there is an obvious discrepancy. }
\label{fig:actthrtwo}
\end{figure}

Different from the volume complexity, which needs careful treatment,  the action complexity of conical defect can be directly obtained from the previous results by setting $\kappa=+1$ in the AdS C-metric \eqref{rot-qubtz}. As shown in Appendix A, the caustics of regularized WdW patch is absent in the black hole case, it is expected to be absent in the conical case as well. 
By direct computation, we find that the late-time growth rate of the action complexity with $\kappa=1$ is non-vanishing only when the horizons exist, i.e. \eqref{bound:a} is satisfied. In this case, the late time derivative admits a semiclassical expansion similar to \eqref{expand-AC}, which is
\be\label{expand-AC-conical}
\lim_{\bar t\to\infty}\frac{dI}{d\bar t}=\frac{\Delta^2\sqrt{\mu^2-4a_0^2}x_1}{2G_3(1+a_0^2x_1^4)}(1+\mathcal{O}(g_{\mathrm{eff}}))
\ee
where $a_0=a/\ell$ and the above expansion is valid for $\frac{\mu}{2}>a_0$.
In Fig. \ref{fig:acttwotwo2}, we compare the late-time growth rate of the action complexity for the rotating quBTZ with $a\gg \ell$ with the one of BTZ, as well as the thermodynamic quantity $TS_{gen}$. Obviously they are in good agreement when $G_3M$ is not small. However, the deviation in the small $G_3M$ limit is significant. The reason is that in this limit, $x_1 \to 0$,  the expansion  \eqref{expand-AC}, which is valid only if $x_1\gg (\ell/\ell_3)^{1/3}$, does not hold. Actually, if we do expansion of \eqref{expand-AC} in the limit $x_1 \to 0$, the late-time slope of action complexity is finite and independent of $a$ 
\be\label{x1:0}
\left. \lim_{\bar t\to\infty}\frac{dI}{d\bar t}\right|_{x_1\to0}=\frac{2\sqrt{1+g_{\mathrm{eff}}^2}}{9G_3}.
\ee
{The same result can be read from the $x_1\to0$ limit of \eqref{expand-AC-conical}.}

Besides $x_1\to0$ limit, there exists another nontrivial limit $a\to0$ which also violates the validity of the expansion \eqref{expand-AC}. In particular, \eqref{expand-AC} fails to be true for $a\ll \ell$, and we need to discuss such case separately.  For simplicity, we simply take $a=g_{\mathrm{eff}}\ell$.  In this case, the inner and outer horizon radius of the quantum-corrected rotating black hole with $\kappa=-1$ are approximated by
\be\label{alorder}  \lim_{\ell\to 0}r_{+}= \ell_{3}+\frac{\mu}{2}\ell, \quad \quad \lim_{l\to 0}r_{-}= \frac{1}{\mu\ell_3^2}\ell^3, \ee 
and for quantum-corrected rotating conical defects with $\kappa=1$,
\be \lim_{\ell\to 0}r_{+}= \mu \ell, \quad \quad \lim_{\ell\to 0}r_{-}= \frac{1}{\mu\ell_3^2}\ell^3.  \ee
Consequently the late-time behavior of the action complexity and thermodynamic quantities $TS_{gen}$ are of the following forms up to the leading order of $\ell$ 
\be
M>0 (\kappa=-1): \hspace{2ex} \lim_{\substack{\ell\to 0\\\bar t\to\infty}} \frac{dI}{d \bar{t}}= \frac{(1+x_1^2) \Delta^2 \ell}{G_{4} x_{1}^2}, \hspace{3ex} \lim_{\ell\to 0}  \text{TS}_{gen}=\frac{\Delta^2 \ell}{2 G_{4}},\ee\be
M<0 (\kappa=+1):\hspace{2ex}  \lim_{\substack{\ell\to 0\\\bar t\to\infty}} \frac{dI}{d \bar{t}}= \frac{(1-x_1^2) \Delta^2 \ell }{G_{4} x_{1}^2}, \hspace{3ex} \lim_{\ell\to 0}  \text{TS}_{gen}=\frac{\Delta^2 \ell}{2 G_{4}}.
\ee
Recalling that $G_4\sim 2G_3\ell$, we can see that both the late-time slope of the action complexity and $TS_{gen}$ are of order $\mathcal{O}(\ell^0)$ in the late time, but the ratio between them is a function of $x_1$, or equivalently a function of $M$, rather than a constant. In Fig. \ref{fig:actthrtwo}, we compare the late-time slop of the action complexity for the rotating quBTZ with the one of BTZ, as well as $TS_{gen}$, in the case of $a=g_{\mathrm{eff}}\ell$. As shown in the figure, the correction from the quantum effect is significant.

\subsection{Discontinuity of non-rotating limit}

In this section, we study why the late-time slope of the action complexity of non-rotating quBTZ cannot be obtained by taking the $a\to0$ limit.  In order to explore the limiting behavior of the action, we show here the out and inner horizon radius in the $a \to 0$ limit (we can see that  \eqref{alorder} is a special case of the following general formula when $a=g_{\mathrm{eff}}\ell$),
\be \label{rpml}
  r_{+}=l_{3}+\frac{\mu \ell}{2}+ a^2 \( \frac{\mu \ell}{\ell_{3}^{2}}-\frac{1}{2 \ell_3} \) , \quad r_{-}=\frac{a^2}{\mu \ell}.
\ee 
Note that when $a \neq 0$ we always have two real positive roots, i.e., two horizons including the inner one $r_{-}$ and outer one $r_{+}$. However when $a=0$ we only have one positive solution. From $\eqref{rpml}$ we can see easily that \be \label{lima} r_{+}|_{a\to 0}= \ell_{3}+\frac{\mu \ell}{2}= r_{s+}, \quad r_{-}|_{a\to 0}=0 \ee
where $r_{s+}$ is the location of the horizon in static quBTZ case.  Also for the rotating black holes, the time-dependent actions include the bulk action, the joint action at  both the future- and past-meeting points. 
 While in the non-rotating quBTZ black hole case, the nontrivial time-dependent parts are the bulk action, the joint action at the past-meeting point and the Gibbons-Hawking-York action at the cutoff surface near the future singularity.   
Let us now compute the non-rotating limit of the above mentioned terms one by one and compare them with the ones in the non-rotating case. 

\paragraph{Bulk action}
Due to the limiting behavior $\eqref{lima}$, the non-rotating limit of the bulk action becomes
\begin{equation}
   \lim_{a\to0} \frac{dI_{bulk}}{d\bar t}=-\frac{\Delta^2 r_{{s+}}^2\ell}{2G_4\ell_4^2}\Big(1-\frac{\ell^2}{(\ell+x_1r_{{s+}})^2}\Big)+\frac{\Delta r_{{s+}}^2}{2G_4\ell}.
\end{equation}
This result agrees with the bulk result of the non-rotating quBTZ black hole which is given by (4.26) in \cite{Emparan:2021hyr}.

\paragraph{Joint action at past}
Similarly, for the joint action at the past-meeting point, the non-rotating limit is also {easily taken}. Setting $a=0$ in \eqref{timedepjoint} with $i=2$, we get
\begin{equation}
     \lim_{a\to0}  \frac{dI^2_{jnt}}{d\bar t}=\frac{\ell x_1r_+^2\Delta^2H'(r_+)}{2G_4(\ell+x_1r_+)}=\frac{\ell x_1\Delta^2 r_+({4 r_{s+}^3-} \mu \ell \ell_3^2-2 {\ell_{3}^2} r_{{s+} })}{2G_4\ell_3^2(\ell+x_1r_+)}.
\end{equation}
where we have used $H(r_-)=0$ to remove the parameter $a$ in the last equality. Again, this result matches with the one in the non-rotating case.

\paragraph{Joint action at future}
Performing the same steps as above and  setting $i=1$ in \eqref{timedepjoint}, we get similar expression 
\begin{equation}
     \frac{dI_{jnt}^1}{d\bar t}=-\frac{\ell x_1r^2_-\Delta^2 H'}{2G_4(\ell+x_1r_-)}=\frac{\ell x_1\Delta^2(-4r_-^3+\ell_3^2(2r_-+\mu\ell))}{2G_4\ell_3^2(\ell+x_1r_-)},
\end{equation}
whose  non-rotating limit is given by setting $r_-=0$,
\be\label{jnt:a=0}
\lim_{a\to0} \frac{dI_{jnt}^1}{d\bar t}=\frac{\Delta^2x_1\mu\ell}{2G_4}.
\ee
For the non-rotating case, the only term remained to have non-trivial time dependence is the GHY term. Its time derivative is given by
\begin{equation}
    \frac{dI_{GHY}}{dt}=\frac{3\Delta^{{2}} x_1\mu\ell}{2G_4}
\end{equation}
which is different from  \eqref{jnt:a=0}. Due to this difference, our result for the rotating quBTZ is very different from that of the non-rotating quBTZ, and the latter cannot be obtained by taking a non-rotating limit.

We find that the mismatch arises from the joint term at the future-meeting point.  We note that as long as $a\neq0$ and no matter how small it is, there will always be a joint surface at the future with some radial coordinate $r_{m_1}$ even at very late time. However, for the non-rotating case, there is a critical time after which the future null surface will touch the singularity at $r=0$. Therefore, the $a\neq 0$ case and the $a=0$ case are essentially different, and the subtleties may arise when we take $r_{m_1}\to r_-$ and  $a\to0$ at the same time.
This is indeed the case.
At very late time, the radial coordinate  of such joint surface approaches $r_-$, which becomes 0 in the non-rotating limit.  
 More precisely, on one hand, the future joint surface approaches to the inner horizon $r_-$ in an exponential way as the time increases, 
\be
r_{m_1}=r_-+Ae^{-4\pi T_-\bar t}
\ee
with $A$ being some factor independent of time and $T_-$ being the temperature defined on the inner horizon.
On the other hand, we have $r_-=O(a^s)$ with  $s$ being some positive constant depending on the blackening factor. Therefore, to make $r_{m_1}\to r_-$ precise, the time $\bar t$ needs to be large enough so that
\be
a^s\gg A e^{-4\pi T_-\bar t},
\ee
which fails to be true if we take the non-rotating limit exactly. According to the above discussions, in order for the approximation $r_{m_1}\to r_-$ 
valid, the angular velocity $a$ cannot be strictly taken to be zero. In other words there has to be an inner horizon. 
 
 In short,  the action complexity of the static quBTZ black holes cannot be reproduced by taking the nonrotating limit $a\to0$ on the one of the rotating black holes.
This discontinuity also appears in higher dimensional rotating black holes, such as Kerr-AdS$_4$ black hole and rotating AdS$_5$ black hole.  We have checked explicitly this singular limit, and leave the details to the appendix C. In principle, one can check this point for other rotating black holes, say Myers-Perry black holes,  and we expect the similar discontinuity.

\section{Conclusion and Discussions}

In this work, we studied holographic complexity for rotating BTZ black holes with quantum corrections from the bulk fields, in the framework of double holography. In particular, we focused on the late-time growth rates of the volume complexity and action complexity, and showed their agreement in most of parameter space at the leading classical order up to a factor 2, but found different quantum corrections.  
For the volume complexity, the leading quantum correction purely comes from  $\delta\text{Vol}(\Sigma_{\bar t})$, originating from semiclassical backreaction on the  geometry. The  contribution from the bulk fields $C_V^{bulk}(|\phi\rangle)$ vanishes to leading order in $g_{\mathrm{eff}}$ which is the same as the static case. 

For the action complexity, if the mass and the spin of BTZ is not small, or more precisely the expansion \eqref{expand-AC} holds, it could be formally expanded in orders of $\ell$ to separate the quantum correction from the classical contribution. The picture is very different from the static case. First of all,  the Wheeler-de Witt (WdW)
patch in computing action complexity for the rotating black hole does not touch the black
hole singularity such that the possible strong quantum effect near the singularity is absent. As a result, the semiclassical expansion of the complexity is feasible. In fact, the leading-order result is in perfect match with the one of classical BTZ, and the quantum correction is not vanishing, in sharp contrast with the static case. 

However, the story is very different when the mass or the spin of BTZ is small. In this case,  different from the volume complexity, the action complexity could receive significant quantum correction of order $\mathcal{O}(\ell^0)$. This is somehow in accord with the picture in static quBTZ. An intuitive interpretation is that even though the WdW patch cannot touch the spacetime singularity, the inner horizon radius is of order $\mathcal{\ell}$, not far from the singularity,  such that the quantum correction there could be significant.

Furthermore, we compared the late-time growth rates of the complexities with the thermodynamic quantity $TS_{gen}$, and found that they are proportional to each other quite well in most of parameter space. More precisely, the late-time behavior of the volume complexity is always well approximated by $TS_{gen}$ as we vary the mass $M$ or angular parameter $a$ of the black hole. However, the late-time growth rate of the action complexity 
shows a large deviation from $2TS_{gen}$ when $M\to0$ or $a\ll \ell$, which suggests that there are  significant quantum corrections.

Another remarkable point is that from our study the nonrotating limit of the late-time growth rate of the action complexity is singular. We found that the joint term in the action is discontinuous in the limit. Moreover we noticed the existence of the same kind of singular limit in the study of the action complexity for the Kerr-AdS spacetimes. 

Although the computations of action complexity are essentially different for the black holes with and without inner horizon, this is not sufficient to conclude that their complexity cannot be related by a limiting procedure, at least numerically. One relatively simple example is the BTZ black holes. It is easy to check that the joint action at future in rotating BTZ black holes agrees with the GHY action near the singularity in non-rotating BTZ black holes when we take the limit $a\to0$. Another example is the charged black holes.
 The Penrose diagram for a charged black hole is similar to a rotating black hole and in particular, it has both inner horizon and outer horizon. However, by direct calculation, the late-time slope of the action complexity is continuous under the chargeless limit. Technically, this is because the Maxwell action contributes to the late-time derivative of the bulk action, which is nonzero even if we take the chargeless limit. It is this nonzero term that compensate for the difference between the future-joint term 
and the GHY term to make the chargeless limit continuous. We  give the details to illustrate this point in the appendix C.

It would be interesting to explore further this discontinuity appearing in the rotating black holes. For example, one can ask if this requires one to modify the definition of the action complexity  to make the limit continuous, and if not, what is the interpretation for such discontinuity on the field theory side. One possible direction is to consider the generalized version of the action complexity proposed in \cite{Belin:2022xmt}. 

\section*{Acknowledgments}
The work is in part supported by NSFC Grant  No. 12275004, 11735001.

\appendix

\section{Absence of caustics}

The caustic point is signalled by the zero point of the determinant $\gamma$ of the induced metric
on the null boundary of the WdW patch, which is given by
\begin{equation}\label{det}
    \sqrt{\gamma}=\frac{\ell^2(1-\tilde a^2)\Delta^2 sPQ}{(\ell+xr)^2}.
\end{equation}
The presence of the caustics will cause difficulty in calculating the action complexity. 
To show that the null surface is free from caustics, we need to show that each factor $s,P$ and $Q$ is nonzero on the null surface. Similar discussion about the caustics in the Kerr-AdS spacetime has been given 
in \cite{AlBalushi:2019obu}, and we follow their procedure to solve \eqref{nullWDW} as the first step. As the function $\xi$ is auxiliary, we would like to find a solution that is independent of $\zeta$.  We first assume that $\zeta$ is now a function of $r$ and $x$. In this case, we have $r^*=\rho(r,x,\zeta)$ where
\be\label{drho}
d\rho=\frac{Q}{\Delta_r}dr+\frac{P}{\Delta_x}dx+\frac{a^2}{2(1-\tilde a^2)^2\Delta^2}F d\zeta,
\ee
where  { $F$ is the partial derivative \be \p_\zeta\rho(r,x,\zeta)= \frac{a^2}{2(1-\tilde a^2)^2\Delta^2} F(r,x,\zeta), \ee} and also we have 
\begin{equation}\label{def-F}
    F {(r,x,\zeta)}=\int^\infty_r\frac{dr'}{Q {(r',\zeta)}}+\int^x_0\frac{dx'}{P {(x',\zeta)}}+g(\zeta)
\end{equation}
with $g(\zeta)$ being an arbitrary integration function {which is important in determining the chosen of null hypersurfaces.} The condition \eqref{dr*} implies that 
\be	\label{eqn-F}
F(r,x,\zeta)=0
\ee
which fixes the {functional form of} $\zeta$ on $(r,x)$ for any given choice of the function $g(\zeta)$. The explicit form of the general solution of \eqref{drho} is then
\be\label{rho}
\rho(r,\theta,\zeta)=\int^r_0\frac{Q {(r',\zeta)}}{\Delta_{r{'}}}dr+\int^x_0\frac{P{(x',\zeta)}}{\Delta_{x{'}}}dx+\frac{a^2}{2(1-\tilde a^2)^2\Delta^2}\int_0^\zeta g(\zeta')d\zeta'
\ee
Note that for a given $g(\zeta)$ appearing in \eqref{def-F} and \eqref{rho}, the integrals  are performed assuming that $\zeta$ is a constant. Then \eqref{eqn-F} is used to solve for $\zeta(r,x)$, which in turn is substituted into the result obtained upon integrating \eqref{rho}. In the end, $r^*(r,x)=\rho(r,x,\zeta(r,x))$ can be obtained.

Next we would like to show that there exists a choice of $g(\xi)$ which makes $s,P$ and $Q$ not vanishing on the null surface. Before we do that, let us first review briefly the case of Kerr-AdS, and compare it with the case at hand.  
In the  case of Kerr-AdS spacetime discussed in \cite{AlBalushi:2019obu}, the exterior derivative of $r^*$ has the same form as \eqref{dr*} with $P,Q,\Delta_r,\Delta_x$ being also quartic polynomials of their respective arguments, but the coefficients are different from the ones in our case. It has been shown in an analytic way \cite{AlBalushi:2019obu} that there is no caustics for the quasi-spherical light cones in Kerr-(A)dS spacetimes. The strategy of the proof in \cite{AlBalushi:2019obu} is firstly show that in a particular $m\to 0$ limit the metric  reduces to the vacuum AdS spacetime, and a specific WdW patch is free from caustics;  secondly, show that each factor that makes up the determinant  of the intrinsic metric of the  null hypersurfaces are all increasing functions of $m$ in the in-going $r$ direction, so that each factor converges less rapidly than $m=0$ case and remains nonzero for $m>0$. The essential point for the possibility to study $m=0$  analytically is that the functions $Q^2$ and $P^2$ becomes even functions of their respective arguments so that the roots of these quartic polynomials come in pairs. In such circumstance, 
the authors in \cite{AlBalushi:2019obu}  can explicitly engineer a function $g(\zeta)$ such that the solution to \eqref{eqn-F}, $\zeta(r,x)$ and the null surface $r^*(r.x)$ can be directly solved.
In our case, however, the integrand cannot be simplified under any sensible limit and therefore we cannot prove the absence of caustics in the same way. Nevertheless we will give an argument that   our WdW patch has no caustics as well. More precisely. we will show that there exists a choice of $g(\zeta)$ such that  the solution $\zeta(r,x)$ would make $s, P$ and $Q$ nonzero on the whole null surface, and the corresponding null surface satisfies  the required boundary conditions as well.
 
Firstly, it is easy to see that for $P^2$  to be positive at any point in the bulk, $\zeta$ needs  a lower bound 
\begin{equation} \label{lower-zeta}
    \zeta (r,x) > \frac{(x^2-x_1^2)^2}{\Delta_x}\geq x_1^4 ,
\end{equation}
where we have used the fact  that $\frac{(x^2-x_1^2)^2}{\Delta_x}$ is decreasing function of $x$ so that it has maximum at $x=0$ with the value $x_1^4$.
Similarly for $Q^2$ to remain positive, $\zeta$ needs an upper bound that depends on $r$,
\begin{equation}
    \zeta<\frac{(r^2+a^2x_1^2)^2}{a^2\Delta_r},\quad \mathrm{for}~\Delta_r\geq0 .
\end{equation}
When $\Delta_r<0$, $Q^2$ is always positive as long as $\zeta>0$. The upper bound $ \frac{(r^2+a^2x_1^2)^2}{a^2\Delta_r}$ is also a decreasing function of $r$ which reaches its minimal value $\frac{\ell_3^2}{a^2}$ when $r\to\infty$ and its maximal value $\infty$ when $r\to r_+$.

To get rid of the caustic point, $\zeta$ is required to satisfy
\begin{itemize}
    \item $x_1^4<\zeta<\frac{\ell_3^2}{a^2}$ so that both $Q$ and $P$ are positive.
    \item $\zeta$ is a decreasing function of $r$ for fixed $x$ so that $s$ is always positive.
\end{itemize}
In the following, we will argue that it is possible to make the solution $\zeta(r,x)$ to \eqref{eqn-F} satisfy $x_1^4<\zeta(r,x)<\ell_3^2/a^2$ when $g(\zeta)$ is chosen properly.

Using \eqref{def-F} and \eqref{eqn-F}, we find
\begin{equation}\label{partial-zeta}
    \begin{aligned}
        &\frac{\partial\zeta}{\partial r}= \frac{1}{Q}\Big[g'(\zeta)+\frac{a^2 }{2(1-\tilde a^2)^2 {\Delta^2} }\int^\infty_r\frac{\Delta_r dr'}{Q^3}-\frac{a^2 }{2(1-\tilde a^2)^2 {\Delta^2} }\int_0^x\frac{\Delta_x dx'}{P^3}\Big]^{-1},\\
        &\frac{\partial\zeta}{\partial x}={-} \frac{1}{P}\Big[g'(\zeta)+\frac{a^2 }{2(1-\tilde a^2)^2 {\Delta^2} }\int^ \infty_r\frac{\Delta_r dr'}{Q^3}-\frac{a^2 }{2(1-\tilde a^2)^2 {\Delta^2} }\int_0^x\frac{\Delta_x dx'}{P^3}\Big]^{-1}.
    \end{aligned}
\end{equation}
This is just \eqref{dzeta} with
\begin{equation}
   {-} s=g'(\zeta)+\frac{a^2 }{2(1-\tilde a^2)^2 {\Delta^2} }\int^\infty_r\frac{\Delta_r dr'}{Q^3}-\frac{a^2 }{2(1-\tilde a^2)^2 {\Delta^2} }\int_0^x\frac{\Delta_x dx'}{P^3}.
\end{equation}
We study \eqref{eqn-F} at $x=0$
\begin{equation}\label{eqn:x0}
    \int^\infty_r\frac{dr'}{Q {(r',\zeta) }}+g(\zeta)=\mathcal{I}_r(r,\zeta)+g(\zeta)=0.
\end{equation}
Given the function $g(\zeta)$, we get the relation between $r$ and $\zeta$ which we denote as $\zeta=\zeta_0(r)$, i.e.
\begin{equation}
    \lim_{x\to0}\zeta(r,x)=\zeta_0(r).
\end{equation}
 Requiring $x_1^4<\zeta<\frac{\ell_3^2}{a^2}$, $\mathcal I_r(r,\zeta)$ is a decreasing function of $r>r_+$ approaching to 0 when $r\to\infty$, and an increasing function of $\zeta$. Choosing $\zeta_a,\zeta_b$ such that
\be
x_1^4<\zeta_a<\zeta_b<\frac{\ell_3^2}{a^2}.
\ee
Then within the region $\zeta\in[\zeta_a,\zeta_b],r\in[r_+,\infty)$, the integral $\mathcal I_r(\zeta,r)$ is bounded by
\be
0<\mathcal I_r(\zeta,r)\leq \mathcal I_r(\zeta_b,r_+).
\ee
We choose $g(\zeta)$ to be a smooth decreasing function of $\zeta$ and satisfy
\be
-g(\zeta)\in[0,I_r(\zeta_b,r_+)],\quad \zeta\in[\zeta_a,\zeta_b].
\ee
We require that $\zeta_a=\zeta_b-\varepsilon$
where $\varepsilon$ is an infinitely small positive number. Such a choice of $g$ implies the existence of a solution $\zeta_0(r)$ such that
\be
\zeta_0(r_+)=\zeta_b,\quad \zeta_0(\infty)=\zeta_b-\varepsilon
\ee
as well as $g'(\zeta)\sim {-} \mathcal{O}(\varepsilon^{-1})$ for $\zeta\in[\zeta_b-\varepsilon,\zeta_b]$, which means that $g(\zeta)$ changes very rapidly. We can extend  this property to a slightly larger region without affecting the existence of solution $\zeta_0(r)$
\be
g'(\zeta)\sim {-}\mathcal{O}(\varepsilon^{-1}),\quad \zeta\in[\zeta_b-\varepsilon,\tilde\zeta_b]
\ee
with $\tilde\zeta_b-\zeta_b=\mathcal{O}(\varepsilon)$ and $\tilde\zeta_b<\ell_3^2/a^2$.
We view such determined function $\zeta_0$ as a boundary condition for $\zeta(r,x)$ at $x=0$
 and then study its evolution into the region $0<x\leq x_1$. We pick up a constant $r_0$ slice then we have
 \be
 \zeta(r_0,x_1)=\zeta_0(r_0)+\int_0^{x_1} dx\frac{\partial\zeta(r_0,x)}{\partial x}>\zeta_0(r_0)
 \ee
 where the derivative of $\zeta$ with respect to $x$ is given by the second line of \eqref{partial-zeta}. 
 Since $x_1$ is finite compared to large $1/\varepsilon$ and $g'(\zeta)\sim \mathcal{O}(\varepsilon)$, it is possible to choose $\tilde\zeta_b$ such that $\zeta(r_0,x_1)\leq \tilde\zeta_b$ for any $r_0$. Consequently, the solution $\zeta=\zeta(r,x)$ obtained by solving  \eqref{eqn-F} satisfies
 \be
 \zeta(r,x)\in[\zeta_b-\varepsilon,\tilde\zeta_b].
 \ee
 This guarantees that both $P^2$ and $Q^2$ remain positive as well as that $g'(\zeta)$ dominates the partial derivatives of $\zeta$ so that $s$ is also positive. Therefore, the null surface is free from caustics according to our construction. The key property that makes the argument possible is $x_1^4<\ell_3^2/a^2$, or equivalently $\tilde a<1$.

 Up to now, we have proved the existence of a choice of $g(\zeta)$ such that the null surface is free from caustics. We also need to check that the corresponding null surface satisfies the required boundary conditions.
In our setup, the boundary conditions consist of two sets. The first one is the usual requirement that
the null surface starts from the time slice at the constant $r=r_{\infty}$ cutoff surface. Note that the null surface is parametrized by the equation $t\pm r^*=$constant. The boundary condition is ensured  as long as $r^*$ satisfies $r^*\sim \lambda r$ when $r\to\infty$, with the coefficient $\lambda$ being some constant that is independent of $x$. This is indeed true for our chosen $g(\zeta)$ it can be seen as follows. Note that $r^*$ is obtained by integrating \eqref{dr*}
\be
r^*=\int^r_0\frac{Q}{\Delta_r}dr+\int^x_0\frac{P}{\Delta_x}dx
\ee
where  we need to substitute the solution $\zeta=\zeta(r,x)$ into the integrand before doing the integration. It is easy to see that the integral is of order $r$ since the integrand of the first term is finite when $r\to\infty$. The fact that the coefficient $r^*/r$ is independent of $x$ for large $r$ can be seen by taking the derivative of $r^*$ with respect to $x$. It is easy to check that the derivative is finite at large $r$ so that the coefficient is a constant. These suffice to prove that the null surface starts from the constant time slice at the asymptotic boundary.

The second set of boundary condition comes from the gluing two identical copies of spacetimes along the brane $x=0$. We expect that the WdW patch is connected across the brane. This implies the boundary condition that the normal vector to the null boundary of the WdW patch should be tangent to the brane as well. This can be realized by requiring that $\zeta_b-\varepsilon=x_1^4,\varepsilon\to0$. Note that although this leads to $P\to0$ when $x=0$, it is a set of measure zero and does not affect  the action complexity. Actually, this is similar to the case of Kerr-AdS$_4$.

\section{Complexity in old coordinates}
The rotating AdS C-metric is of the form
\begin{equation}
    \begin{aligned}
       ds^2=\frac{\ell^2}{(\ell+xr)^2}\Big[&-\frac{H(r)}{\Sigma(x,r)}(dt+ax^2d\phi)^2+\frac{\Sigma(x,r)}{H(r)}dr^2\\
    &+r^2\Big(\frac{\Sigma(x,r)}{G(x)}dx^2+\frac{G(x)}{\Sigma(x,r)}(d\phi-\frac{a}{r^2}dt)^2\Big)   \Big]
    \end{aligned}
\end{equation}
where
\begin{equation}
    \begin{aligned}
       &H(r)=\frac{r^2}{\ell_3^2}+\kappa-\frac{\mu\ell}{r}+\frac{a^2}{r^2},\\
       &G(x)=1-\kappa x^2-\mu x^3+\frac{a^2}{\ell_3^2}x^4,\\
       &\Sigma(x,r)=1+\frac{a^2x^2}{r^2}.
    \end{aligned}
    \end{equation}
    Consider a general linear mixing between $t$ and $\phi$ coordinates\footnote{We ignored an overall factor $\Delta$ compared to \eqref{redf} which does not affect the analysis.}
    \begin{equation}
        t\to t+k_1\phi,\quad \phi\to \phi+k_2t.
    \end{equation}
    The canonical coordinates $\bar t$ and $\bar\phi$ correspond to the choice $k_1=-\tilde a\ell_3,k_2=-\frac{\tilde a}{\ell_3}$.
The metric components then become
\begin{equation}
    \begin{aligned}
        &\frac{(\ell+xr)^2}{\ell^2}g_{tt}=-\frac{H(1+k_2ax^2)^2}{\Sigma}+\frac{r^2G}{\Sigma}(k_2-\frac{a}{r^2})^2,\\
        &\frac{(\ell+xr)^2}{\ell^2}g_{t\phi}=-\frac{H}{\Sigma}(1+k_2ax^2)(k_1+ax^2)+\frac{r^2
        G}{\Sigma}(k_2-\frac{a}{r^2})(1-\frac{ak_1}{r^2}),\\
        &\frac{(\ell+xr)^2}{\ell^2}g_{\phi\phi}=-\frac{H}{\Sigma}(k_1+ax^2)^2+\frac{r^2G}{\Sigma}(1-\frac{ak_1}{r^2})^2.
    \end{aligned}
\end{equation}
For this transformed metric,  we have
\begin{equation}
    g^{tt}=\frac{g_{\phi\phi}}{g_{tt}g_{\phi\phi}-g_{t\phi}^2}=\frac{(\ell+xr)^2}{\ell^2(1-k_1k_2)^2}\Big[\frac{(ax^2+k_1)^2}{r^2G\Sigma}-\frac{(1-ak_1/r^2)^2}{H\Sigma}\Big].
\end{equation}
The null condition 
\begin{equation}
    g^{tt}+g^{rr}(\partial_rr^*)^2+g^{xx}(\partial_xr^*)^2=0
\end{equation}
 can be rewritten as
\begin{equation}\label{null}
   r^2 H(\partial_rr^*)^2+G(\partial_xr^*)^2=\frac{1}{(1-k_1k_2)^2}\Big[\frac{(r^2-ak_1)^2}{r^2H}-\frac{(ax^2+k_1)^2}{G}\Big].
\end{equation}
The left-hand side is non-negative for $r>r_+$. Since $G(x_1)=0$, requiring the positivity of the right-hand side of \eqref{null} implies
\begin{equation}
    ax^2_1+k_1=0~\Rightarrow~ k_1=-ax_1^2=-\tilde a\ell_3.
\end{equation}
So $k_1$ is uniquely determined for consistency. 
Otherwise, the null condition is broken when $x$ is close to $x_1$ which suggests that the null surface does not exist there. In other words, if $k_1\neq -\tilde a\ell_3$, the WdW patch formed by a constant time slice is ill defined.

\section{Non-rotating limit vs chargeless limit}

From the discussion on the rotating quBTZ black hole in the main text,  we have  seen that in the non-rotating limit, the late-time slope of  the bulk action and the joint action at the past intersecting surface could reduce to the ones of non-rotating black hole. But the joint action at the future intersecting surface fails to reduce to the GHY term for the non-rotating black hole. In this appendix we discuss the same non-rotating limit for the action complexity of Kerr-AdS$_5$ black hole. This case has the advantage that the null surface is determined by an ordinary differential equation which allows us to study in more detail. For comparison, we investigate the chargeless limit of the action complexity of charged black hole, which has two horizons, as well. 

\subsection{5d rotating black holes in AdS}


For the  rotating black hole in AdS$_5$, it has the following metric\cite{Castro:2018ffi}
\be\label{5d-met}
ds^2_5=	-\frac{1}{\Xi}\Delta(r)e^{U_2-U_1}dt^2+\frac{r^2dr^2}{(r^2+a^2)\Delta(r)}+e^{-U_1}d\Omega_2^2+e^{-U_2}((\sigma^3+A)^2
\ee
where
\be\ba
e^{-U_2}&=\frac{r^2+a^2}{4\Xi}+\frac{ma^2}{2\Xi^2(r^2+a^2)},	\\
e^{-U_1}&=\frac{r^2+a^2}{4\Xi},\\
A&=A_tdt=\frac{a}{2\Xi}\Big(\frac{r^2+a^2}{\ell_5^2}-\frac{2m}{r^2+a^2}\Big)e^{U_2}dt,
\ea
\ee
with
\be
\Xi=1-\frac{a^2}{\ell_5^2}	,\quad \Delta(r)=1+\frac{r^2}{\ell_5^2}-\frac{2mr^2}{	(r^2+a^2)^2}.
\ee
The  angular forms $\sigma^i$ are 
\begin{equation}
\begin{aligned}
&\sigma^1=-\sin\psi d\theta+\cos\psi\sin\theta d\phi,\\
&\sigma^2=\cos\psi d\theta+\sin\psi\sin\theta d\phi,\\
&\sigma^3=d\psi+\cos\theta d\phi,
\end{aligned}\end{equation}
and in the metric there is 
\be
d\Omega_2^2=d\theta^2+\sin^2\theta d\phi^2=(\sigma^1)^2+(\sigma^2)^2.	
\ee
Taking the limit $a\to0$, the metric simply reduces to a Schwarzschild-AdS black hole
\be
\lim_{a\to0}ds_5^2=-\Delta_0(r)dt^2+\frac{dr^2}{\Delta_0(r)}+\frac{r^2}{4}\Big((\sigma^1)^2+(\sigma^2)^2+(\sigma^3)^2\Big)
\ee
where
\be
\Delta_0(r)=\lim_{a\to0}\Delta(r)=1+\frac{r^2}{\ell_5^2}-\frac{2m}{r^2}.
\ee

Due to the fact that the metric components are only the functions of $r$, the null surfaces in  the spacetime described by the metric \eqref{5d-met}  are parametrized by $u=$const. or $v=$const., where
\begin{equation}
v=t+r^*(r),\quad u=t-r^*(r)
\end{equation}
with
\be
\frac{dr^*}{dr}=\Big(-g^{tt}/g^{rr}\Big)^{1/2}=\Big(\frac{\Xi r^2e^{-U_2+U_1}}{r^2+a^2}\Big)^{1/2}\Delta^{-1}:=F(r).
\ee
We again set $t_L=t_R=t$ and denote the intersecting surface at future by its radial coordinate $r_c$. Then we have
\be
F\frac{dr_c}{dt}=1,\quad\lim_{t\to\infty}r_c=r_-
\ee
 with $r_-$ being the inner horizon. To evaluate the joint action, we use the following outward-directed null normal co-vectors:
\be
k_v=\alpha_1dv=\alpha_1(1,F,0,0,0),\quad k_u=-\alpha_2du=\alpha_1(-1,F,0,0,0)
\ee
with $\alpha_1,\alpha_2$ being two positive constants. The inner product is then given by
\be
|k_u\cdot k_v|=|2\alpha_1\alpha_2\Xi e^{-U_2+U_1}\Delta^{-1}|.
\ee
The joint surface has constant $t$ and $r$ so we have
\be
\int d^3y\sqrt{\sigma}=\int d\theta d\phi d\psi e^{-U_1-U_2/2}\sin\theta.
\ee
Putting all things together, the joint action is given by   
\be
I_{jnt}=\frac{1}{8\pi G_5}\int d\theta d\phi d\psi e^{-U_1-U_2/2}\sin\theta\log\Big|\alpha_1\alpha_2\Xi e^{-U_2+U_1}\Delta^{-1}\Big|.
\ee
Its late-time derivative is given by 
\be
\lim_{t\to \infty}\frac{dI_{jnt}}{dt}=-\frac{1}{8\pi G_5}\int d\theta d\phi d\psi e^{-U_1-U_2/2}\sin\theta\frac{\Delta'}{\Delta}F^{-1}\Big|_{r=r_-}.
\ee
If we further take the limit $a\to0$, we have $r_-\sim\frac{a^2}{\sqrt{2m}}$ which implies
\be
\lim_{a\to 0}\lim_{t\to\infty}\frac{dI_{jnt}}{dt}=\frac{1}{8\pi G_5}\int d\theta d\phi d\psi\sin\theta\frac{m}{2}=\frac{\pi m}{4G_5}.\label{latetimeKerrAdS5}
\ee

To compare with the above $a\to 0$ limit, let us consider the action complexity of the non-rotating black hole. The possible mismatch stems from the GHY action. In this case, the regulator surface $r=\epsilon_0\to0$ contributes to the GHY action
\be
I_{GHY}=\frac{1}{8\pi G_5}\int  d^4y\sqrt{|h|}K,
\ee
where the integration over $t$ ranges from $-t-r^*(r_c)+r^*(\epsilon_0)$ to $t+r^*(r_c)-r^*(\epsilon_0)$, and the determinant of the induced metric is 
\be
\sqrt{|h|}=\frac{\sqrt{2m}}{8}r^2\sin\theta+O(r^4).
\ee
As the outward-directed unit normal for a constant $r=\epsilon_0$ surface is given by
\be
\mathbf n=-\frac{dr}{\sqrt{-\Delta(\epsilon_0)}}\Big|_{a=0},
\ee
the trace of the extrinsic curvature is 
\be
K=\frac{n_r}{2}	\Big(\Delta'+\frac{6}{r}\Delta\Big)=\frac{2\sqrt{2m}}{r^2}+O(1),
\ee
so we have
\be
\frac{dI_{GHY}}{dt}=\frac{2}{8\pi G_5}\int d\theta d\phi d\psi \sqrt{|h|}K=\frac{\pi m}{2G_5}
\ee
which is twice of \eqref{latetimeKerrAdS5}, the non-rotating limit of the late-time slope of the future-joint action. Thus we see once again the non-rotating limit is singular. 

\subsection{Charged black holes}
In this subsection, we review the growth rate of the action complexity for charged black holes in $(d+1)$ dimensions with $d\geq3$. The charged black holes are the solutions in the Einstein AdS gravity coupled to a Maxwell field with the following action
\be
I=I_{EH}-\frac{1}{4}\int d^{d+1}x\sqrt{-g}F^2.
\ee
The black hole metric  takes the form
\be
ds^2=-f(r)dt^2+\frac{dr^2}{f(r)}+r^2d\Omega^2_{d-1}
\ee
with the blackening factor given by
\be
f(r)=\frac{r^2}{L^2}+1-\frac{\omega^{d-2}}{r^{d-2}}+\frac{q^2}{r^{2(d-2)}}
\ee
where $L$ is the AdS radius,  and $q$ is proportional to the charge of the black hole. The blackening factor has two real roots, $r_+$ and $r_-$, corresponding to the outer horizon and the inner horizon respectively. The gauge potential can be written as
\be
A_t=\frac{1}{2\sqrt{2\pi G_N}}\sqrt{\frac{d-1}{d-2}}\left(\frac{q}{r_+^{d-2}}-\frac{q}{r^{d-2}}\right).
\ee
 The time derivative of the action complexity was computed in \cite{Carmi:2017jqz}, and we summarize the main results below. The contribution from the bulk action, which consists of the Einstein action $I_{EH}$ and the Maxwell action $I_{Max}$, and the joint actions $I_{jnt}^1, I_{jnt}^2$ at future- and past-intersecting surfaces are given by
 \be\ba
 \frac{dI_{EH}}{dt}&=\frac{\Omega_{d-1}}{16\pi G_N}\left(\frac{(d-3)q^2}{r^{d-2}}-\frac{2r^d}{L^2}\right)\big|^{r_2}_{r_1},\\
 \frac{dI_{Max}}{dt}&=\frac{\Omega_{d-1}(1-d)q^2}{16\pi G_Nr^{d-2}}\big|^{r_2}_{r_1},\\
  \frac{dI^1_{jnt}}{dt}&=-\frac{\Omega_{d-1}}{16\pi G_N}\left[(d-1)r^{d-2}f(r)\log\frac{|f(r)|}{\alpha^2}+r^{d-1}f'(r)\right]\big|_{r=r_1},\\
   \frac{dI_{jnt}^2}{dt}&= \frac{\Omega_{d-1}}{16\pi G_N}\left[(d-1)r^{d-2}f(r)\log\frac{|f(r)|}{\alpha^2}+r^{d-1}f'(r)\right]\big|_{r=r_2},
 \ea\ee
where $r_1$ and $r_2$ are the radial coordinates of the future- and past-intersecting surface respectively. In the late-time limit, we have $r_1\to r_-$ and $r_2\to r_+$.
As a result, we have
 \be\ba
\lim_{t\to\infty} \frac{dI_{EH}}{dt}&=\frac{\Omega_{d-1}}{16\pi G_N}\left(\frac{(d-3)q^2}{r^{d-2}}-\frac{2r^d}{L^2}\right)\big|^{r_+}_{r_-},\\
\lim_{t\to\infty} \frac{dI_{Max}}{dt}&=\frac{\Omega_{d-1}(1-d)q^2}{16\pi G_Nr^{d-2}}\big|^{r_+}_{r_-},\\
\lim_{t\to\infty}  \frac{dI^1_{jnt}}{dt}&=-\frac{\Omega_{d-1}}{16\pi G_N}\left[r^{d-1}f'(r)\right]\big|_{r=r_-},\\
\lim_{t\to\infty}   \frac{dI_{jnt}^2}{dt}&= \frac{\Omega_{d-1}}{16\pi G_N}\left[r^{d-1}f'(r)\right]\big|_{r=r_+}.
 \ea\ee
If we further take the chargeless limit $q\to0$ and use the relation
\be
\lim_{q\to0} { r_{-}=\frac{q^{\frac{2}{d-2} } }{w} \to 0
}  , \quad {\lim_{q\to 0} r_{+}=r_{0+}}
\ee
where $r_{0+}$ is the horizon of the chargeless AdS black hole. Then we get
 \be\ba
\lim_{q\to0}\lim_{t\to\infty} \frac{dI_{EH}}{dt}&=\frac{\Omega_{d-1}}{16\pi G_N}\left(-(d-3)w^{d-2}-\frac{2r^d_+}{L^2}\right),\\
\lim_{q\to0}\lim_{t\to\infty} \frac{dI_{Max}}{dt}&=\frac{\Omega_{d-1}(d-1)w^{d-2}}{16\pi G_N},\\
\lim_{q\to0}\lim_{t\to\infty}  \frac{dI^1_{jnt}}{dt}&=\frac{\Omega_{d-1}(d-2)w^{d-2}}{16\pi G_N},\\
\lim_{q\to0}\lim_{t\to\infty}   \frac{dI_{jnt}^2}{dt}&=\frac{\Omega_{d-1}}{16\pi G_N} \left(\frac{2r_+^d}{L^2}+(d-2)w^{d-2}\right).
\ea\ee
As a result, the chargeless limit of the total growth rate at late time is given by
\be
\lim_{q\to0}\lim_{t\to {\infty}}\frac{dI_{tot}}{dt}=\frac{(d-1)\Omega_{d-1}w^{d-2}}{8\pi G_N}=\frac{2M }{\pi}.
\ee
where $M$ is the mass of the black hole. This agrees with the growth rate of the action complexity of $(d+1)$ dimensional Schwarzschild-AdS black hole. Besides, it is  straightforward to check that the chargeless limit of the Einstein action and the past joint action match with  the ones of Schwarzschild-AdS black hole \cite{Carmi:2017jqz}, and more importantly the sum of the Maxwell action and future-joint action reproduces the GHY action in the chargeless limit.

\bibliographystyle{JHEP}
\bibliography{main}
\end{document}
